\documentclass[12pt, floatfix]{iopart}
\usepackage{amsfonts}
\usepackage{graphicx,epsfig}
\usepackage{dcolumn}
\usepackage{bm}
\usepackage{subfigure}
\usepackage{amssymb}
\usepackage{float}
\usepackage{multirow}
\def\lsim{\mathrel {\vcenter {\baselineskip 0pt \kern 0pt \hbox{$<$} \kern 0pt \hbox{$\sim$} }}}
\def\gsim{\mathrel {\vcenter {\baselineskip 0pt \kern 0pt \hbox{$>$} \kern 0pt \hbox{$\sim$} }}}

\newcolumntype{.}{D{.}{.}{4}}
\newcolumntype{p}{D{(}{(}{2}}

\renewcommand{\thesubfigure}{(\roman{subfigure})}
\makeatletter
\renewcommand{\@thesubfigure}{\thesubfigure\space} 
\renewcommand{\p@subfigure}{\thefigure}
\def\bdm{\begin{displaymath}}
\def\edm{\end{displaymath}}

\begin{document}

\title[The Inflationary Parameter Space after WMAP5]{Primordial Black Holes, Eternal Inflation, and the Inflationary Parameter Space after WMAP5}
\author{Hiranya V. Peiris${}^1$ and Richard Easther${}^2$ }
\address{
${}^1$  Institute of Astronomy, University of Cambridge, Cambridge CB3 0HA, U.K.\\
hiranya@ast.cam.ac.uk} 
\address{${}^2$  Department of Physics, Yale University, New Haven,  CT 06520, U.S.A.\\
richard.easther@yale.edu
}

\date{\today}

\begin{abstract}
We consider constraints on inflation driven by a single, minimally coupled scalar field in the light of the WMAP5 dataset, as well as ACBAR and the SuperNova Legacy Survey. We use the Slow Roll Reconstruction algorithm to derive optimal constraints on the inflationary parameter space. The scale dependence in the slope of the scalar spectrum permitted by WMAP5 is large enough to lead to viable models where the small scale perturbations have a substantial amplitude when extrapolated to the end of inflation.   We find that excluding parameter values which would cause the overproduction of primordial black holes or even the onset of eternal inflation leads to potentially significant constraints on the slow roll parameters.  Finally, we present a more sophisticated approach to including priors based on the total duration of inflation, and discuss the resulting restrictions on the inflationary parameter space.

  \end{abstract}

\maketitle

\section{Introduction}

Cosmology has long had the goal of {\em reconstructing\/} the inflationary potential from observational data \cite{Copeland:1993jj,Copeland:1993ie,Turner:1993su,Liddle:1994cr}.  This was originally envisaged as a two-step process. One would first obtain constraints on spectral parameters such as the scalar and tensor spectral indices, $n_s$ and $n_t$, their runnings, and their overall amplitudes. Armed with this information,  one could invert the inflationary  formulae for these variables to obtain the slow roll parameters. These parameters involve derivatives of the potential and would provide a Taylor expansion for the segment of the inflaton potential traversed as  astrophysical perturbations  leave the horizon.  It is now apparent that the optimal approach to reconstruction is to insert the slow roll variables directly into the cosmological parameter set; a process known as {\em Slow Roll Reconstruction} \cite{Peiris:2006sj,Peiris:2006ug,Easther:2006tv,Adshead:2008vn}.  One thus never computes $n_s$ {\em et al.}, but instead fits directly to the slow roll parameters themselves. The flow hierarchy \cite{Hoffman:2000ue,Kinney:2002qn,Easther:2002rw,Kinney:2006qm,Powell:2007gu} automatically accounts for the scale dependence of the slow roll parameters. Reconstruction thus explicitly realizes the hope that the early universe can be used as a ``laboratory'' for high energy physics: the slow roll parameters can be determined by GUT or stringy physics, and we are directly constraining their values with our Markov Chains. 
 
 The flow hierarchy can be solved exactly to yield the underlying inflationary potential as a function of the first $M$ non-zero slow roll parameters   \cite{Liddle:2003py}.  We can thus compute the amount of inflation remaining after a specific mode leaves the horizon, as a function of the slow roll parameters. 
This provides a self-consistency check: fitting to $M$ slow roll parameters carries the tacit assumption that these parameters fully characterize the portion of the inflationary epoch during which observable modes exit the Hubble horizon. Under the assumption that the same parameters characterize the inflaton dynamics till the end of inflation, one obtains even stronger constraints by requiring a sufficiently long inflationary epoch in order to solve the usual cosmological problems. In fact, given the  quality of present-day data, much of the power of Slow Roll Reconstruction derives from priors based on the number of e-folds  \cite{Adshead:2008vn}, especially when we go beyond the first two slow roll parameters. Further, thanks to the analytical solution of the slow roll {\em hierarchy},  our one use of the slow roll {\em approximation\/} is the computation of the  perturbation spectrum. This computation can be performed precisely by numerically solving the mode equations, as pointed out in \cite{Peiris:2006sj,Grivell:1996sr} and implemented in \cite{Lesgourgues:2007aa,Hamann:2008pb}. However, even with the WMAP5 dataset,  the quality of current cosmological data is not high enough to require the use of an exactly computed spectrum, and we do not employ it here, instead using the Stewart-Lyth formulae \cite{Stewart:1993bc} for the power spectra along with the exact solution to the background given by the flow hierarchy.  
 
There are a number of antecedents to Slow Roll Reconstruction. As noted above, the ``inverse problem'' posed by the extraction the inflationary potential from the observed power spectrum was first discussed over 15 years ago.   Leach and collaborators  \cite{Leach:2002ar,Leach:2002dw, Leach:2003us} write the spectral indices as functions of the slow roll parameters at a given pivot, and then constrain the slow roll parameters with data.   If one includes higher order slow roll terms when computing the spectral indices, this approach approximates a full solution of the slow roll hierarchy. However, since the full potential is never computed, one cannot eliminate sets of slow roll parameters for which the number of e-folds is unacceptably small.  Conversely, {\em Monte Carlo Reconstruction\/} \cite{Kinney:2002qn,Easther:2002rw,Kinney:2006qm} specifies an inflationary trajectory, and then uses the flow equations to move to a point some 50 or 60 e-foldings after the observable scale leaves the horizon. However, this process does not provide an unambiguous spectrum when inflation apparently continues indefinitely, and implicitly rules out the possibility of a hybrid transition in models where the potential does naturally lead to the termination of inflation.   This ambiguity is discussed in passing in \cite{Easther:2002rw}, and we will return to it below in Section \ref{sec:bh}.  Moreover, as originally implemented, Monte Carlo reconstruction did not work directly with likelihood functions derived from astrophysical data, and thus did not weight different candidate potentials according to their agreement to astrophysical data.
 
In this paper, we improve the implementation of Slow Roll  Reconstruction by developing a physically realistic e-fold prior \cite{Adshead:2008vn}.   This constraint is a strong function of the $\epsilon$ parameter, which specifies the relative amplitudes of the primordial tensor and scalar spectra. Secondly we present constraints on the inflationary parameter space derived from the recent CMB (Cosmic Microwave Background) data: the 5-year WMAP dataset (WMAP5) \cite{Hinshaw:2008kr, Nolta:2008ih, Dunkley:2008ie, Komatsu:2008hk} and the 2008 ACBAR release (ACBAR) \cite{Reichardt:2008ay}, as well as high redshift type Ia supernovae from the SuperNova Legacy Survey (SNLS) \cite{Astier:2005qq}.  In addition, we show that the constraints on the scale dependence (``running'') of the scalar spectral index derived from WMAP5 data allow models in which the amplitude of the density perturbation approaches unity before the end of inflation.  These scenarios can be ruled out on physical grounds -- either through the overproduction of primordial black holes  \cite{Hawking:1971ei,Carr:1974nx,Green:1997sz,Yokoyama:1999xi,Leach:2000ea,Chongchitnan:2006wx,Zaballa:2006kh,Kohri:2007qn}, or because the inflaton is moving so slowly that its motion would be dominated by its quantum fluctuations, rather than its semi-classical rolling.  The latter situation is equivalent to  eternal inflation \cite{Linde:1986fd}, and inflation cannot terminate coherently over a volume large enough to contain our visible universe.  The parameter space is tightly constrained by excluding scenarios that lead to the overproduction of primordial black holes.  Likewise, a substantial region of the parameter space which permits the onset of eternal inflation is permitted by the WMAP5 data, but most of this region is already excluded by the combination of WMAP5 and SNLS.  This constraint is more important to the WMAP5 dataset than it would have been with WMAP3, which excluded a positive running at the 2$\sigma$ level. The central value for the running obtained from analyses of the WMAP5 dataset is still negative, but the overall preference for a negative running is significantly diminished. 

\section{Slow Roll Reconstruction}

\subsection{Formalism}

For inflation driven by single, minimally coupled scalar field in a spatially flat FRW universe the equations  motion can be written with the field $\phi$ as the independent variable \cite{Lidsey:1995np, Grishchuk:1988,Muslimov:1990be,
Salopek:1990jq,Salopek:1990re,Lidsey:1991zp}
\begin{equation}\label{dotphi}
\dot{\phi}  =  -\frac{m_{\rm{Pl}}^{2}}{4\pi}H'(\phi),
\end{equation}
\begin{equation}\label{Hamilton-Jacobi}
[H'(\phi)]^{2}-\frac{12\pi}{m_{\rm{Pl}}^{2}}H^{2}(\phi) =
-\frac{32\pi^{2}}{m_{\rm{Pl}}^{4}}V(\phi),
\end{equation}
where $m_{\rm Pl} = 1.22 \times 10^{19} $ GeV is the Planck mass. 
The HSR [Hubble slow roll] parameters
$^{\ell}\lambda_{H}$ are defined by the infinite hierarchy of
differential equations \cite{Kinney:2002qn}
\begin{eqnarray}\label{epsphi}
\epsilon(\phi) & \equiv &
\frac{m_{\rm{Pl}}^{2}}{4\pi}\left[\frac{H'(\phi)}{H(\phi)}\right]^{2},\\
^{\ell}\lambda_{H} & \equiv &
\left(\frac{m_{\rm{Pl}}^{2}}{4\pi}\right)^{\ell}\frac{(H')^{\ell-1}}{H^{\ell}}
\frac{d^{\ell+1}H}{d\phi^{(\ell+1)}};\;\ell \geq 1.
\end{eqnarray}
The usual slow roll parameters are $\eta = {}^{1}\lambda_{H}$ and
$\xi = {}^{2}\lambda_{H}$. If we truncate the hierarchy,  so that
${}^{\ell}\lambda_{H} =0 $ for all $\ell > M$ at some $\phi_\star$,
then the ${}^{\ell}\lambda_{H}$ vanish everywhere. When truncated at
order $M$, the hierarchy can be solved explicitly \cite{Liddle:2003py}, allowing us to obtain an exact expression for the potential. Consequently, we can then determine the value of $\phi$ at which inflation ends and the number of e-folds that remain when $\phi = \phi_\star$.   

This formulation includes the implicit assumption that the scalar field dynamics is well-approximated by the truncated hierarchy.  We do not insist that the truncated hierarchy approaches the exact limit as $M\rightarrow \infty$, but only ask that the truncated $H(\phi)$ is an asymptotic expansion of the exact $H(\phi)$ for values of $\phi$ which are astrophysically relevant, with $M$  less than the value at which the asymptotic expansion would begin to diverge.  This covers a very large class of potentials, but excludes ``step'' models where each successive slow roll parameter is larger than its predecessor in a very small range of field values \cite{Adams:2001vc}.    

The scalar and tensor perturbation spectra are given by  \cite{Stewart:1993bc}
\begin{eqnarray}\label{Pr}
P_{\mathcal{R}} & = &
\frac{[1-(2C+1)\epsilon+C\eta]^{2}}{\pi\epsilon}\left.\left(\frac{H}{m_{\rm{Pl}}}\right)^{2}\right|_{k=aH},\\
\label{Ph}P_{h} & = &
[1-(C+1)\epsilon]^{2}\frac{16}{\pi}\left.\left(\frac{H}{m_{\rm{Pl}}}\right)^{2}\right|_{k=aH},
\end{eqnarray}
where $C = -2+\ln 2+\gamma\approx-0.729$ and $\gamma$ is the
Euler-Mascheroni constant. The scale dependence of the spectra then follows from the scale dependence of  $\epsilon$, $\eta$ and $H$.  In our Monte-Carlo Markov Chain (MCMC) analysis, the power spectrum is fixed by setting $A_s = P_{\mathcal{R}}$ at a ``fiducial'' scale  $k_\star$, where we also specify the initial values of the slow roll parameters. We then compute $H_\star$ via equation~(\ref{Hamilton-Jacobi}),  which then fixes the overall scale of inflation.  In practice $A_s$ is effectively fixed by the data, putting tight bounds on $H_\star^2 /\epsilon_\star$, since $\epsilon_\star,\eta_\star \ll 1$.  If $\epsilon_\star$ is very small,   $H_\star$ is also small, lowering $P_{h}$ relative to $P_{\mathcal{R}}$, and reflecting the well-known inflationary consistency condition. We take $k_\star=0.02$~Mpc$^{-1}$, which was the optimal value for WMAP data found in \cite{Peiris:2006ug,Cortes:2007ak}.   Moreover, a useful feature of Slow Roll Reconstruction is that we can post-process our Markov chains using the flow hierarchy to move our constraints to an arbitrary $k_\star$ \cite{Peiris:2006ug}.

\subsection{The $\epsilon$ Parameter and the Duration of Inflation}

 Consider the ``connection equation'' for scales in a universe which inflated, reheated, and passed through matter-radiation equality,
\begin{equation}
\label{connect}
\frac{k}{a_0 H_0} = \frac{a_k H_k}{a_0 H_0} = \exp[-N(k)] \ \frac{a_{\rm end}}{a_{\rm reh}}\frac{a_{\rm reh}}{a_{\rm eq}} \frac{H_k}{H_{\rm eq}} \frac{a_{\rm eq}H_{\rm eq}}{a_0 H_0}.
\end{equation}
Here $H_k$ is the value of the Hubble parameter when the mode with comoving wavenumber $k$ leaves the horizon, a subscript $0$ refers to the present day value, while``eq", ``reh" and ``end" denote values at matter-radiation equality, reheating (i.e. the point at which the universe thermalizes after inflation) and the end of inflation. 
Now assume that the universe is effectively matter dominated between the end of inflation and reheating, so $\rho \propto a^{-3}$. Between reheating and matter-radiation equality, the universe is radiation dominated and $\rho \propto a^{-4}$. Equation~(\ref{connect}) becomes
\begin{eqnarray}
N(k) &=& -\ln\left(\frac{k}{a_0 H_0}\right) + \frac{1}{3}\ln\left(\frac{\rho_{\rm reh}}{\rho_{\rm end}}\right) + \frac{1}{4}\ln\left(\frac{\rho_{\rm eq}}{\rho_{\rm reh}}\right) \nonumber \\ 
& &+ \ln\left(\frac{H_k}{H_{\rm eq}}\right) + \ln\left(\frac{a_{\rm eq}H_{\rm eq}}{a_0 H_0}\right).
\end{eqnarray}
Using the first Friedmann equation in a flat universe, $H \propto \sqrt{\rho}$, along with the usual convention  $a_0 = 1$, we rewrite the last term as an expression involving ${\rho_{\rm eq}}$ and $\rho_{\rm crit}$.  Recall that $a_{\rm eq} =  4.15 \times 10^{-5}/\Omega_m h^2$ \cite{Dodelson:2003ft}, where $\Omega_m$ is the present matter density, and $h$ is the Hubble parameter in units of 100 kms$^{-1}$ Mpc$^{-1}$, and 
$\rho_{\rm eq}/\rho_{\rm crit} = 2\Omega_m/a_{\rm eq}^3$, so
\begin{eqnarray}
\label{leachequiv}
N(k) &=& -\ln\left(\frac{k}{a_0 H_0}\right) + \frac{1}{3}\ln\left(\frac{\rho_{\rm reh}}{\rho_{\rm end}}\right) + \frac{1}{4}\ln\left(\frac{\rho_{\rm eq}}{\rho_{\rm reh}}\right)\nonumber \\ 
& & + \ln\left(\frac{H_k}{H_{\rm eq}}\right) + \ln 220\ \Omega_m h. 
\end{eqnarray}
Up to rounding issues,   this formula is identical to equation~(6) of  \cite{Liddle:2003as}.  During inflation $\rho \approx V(\phi)$, but  there is little to be gained from this substitution here, and we express our free parameters in terms of $H$. We have
\begin{equation}
 H_{\rm eq} = 5.29 \times 10^6\ h^3\ \Omega_m^2\ H_0  
	= 9.24 \times 10^{-55}\ h^4\ \Omega_m^2\ m_{\rm Pl}, 
\end{equation}
recalling that $H_0 = 1.75 \times 10^{-61}\ h\ m_{\rm Pl}$ in natural units.   Substituting for $H_{\rm eq}$ and $\rho_{\rm eq}$, the $\Omega_m$ dependence cancels. After reheating, we can specify the energy density of the universe in terms of  the temperature $T$, but the relationship between $T$ and $H$ depends on the number of degrees of freedom.   Numerically, $H\sim \rho^{1/2}/ {m_{\rm Pl}}$, so while $\rho^{1/4}$ and $H$ have the same units, $H$ is numerically smaller than $\rho^{1/4}$ or $T$.    We  specify $k$ in Mpc$^{-1}$, and with $c=1$,    $H_0 = 100 h/(2.99792 \times 10^5)$ Mpc$^{-1}$, so
\begin{eqnarray}
\label{lnconnect}
N(k) &=& -\ln\left(\frac{k}{{\rm Mpc}^{-1}}\right)+ \frac{1}{6}\ln\left(\frac{H_{\rm reh}}{m_{\rm Pl}}\right) - \frac{2}{3}\ln\left(\frac{H_{\rm end}}{m_{\rm Pl}}\right) + \ln\left(\frac{H_k}{m_{\rm Pl}}\right) \nonumber \\
& & +59.59.
\end{eqnarray}
In practice, $H_k$ and  $H_{\rm end}$ are usually very similar, but reheating need not happen efficiently, so $H_{\rm reh}$ can be far smaller than $H_{\rm end}$.   If the universe is in thermal equilibrium above the neutrino freeze-out temperature (around 1 MeV), the present number density of neutrinos is a function of the number density of photons, which is given by the absolute temperature of the CMB.   By default, the Boltzmann code CAMB ~\cite{Lewis:1999bs} we use to compute the power spectrum assumes  three (almost) massless neutrino species. Thus, to self-consistently compute CMB spectra within our Markov chains, we  need $T_{\rm reh} \gsim 1$ MeV.  The neutrino sector makes a subdominant contribution to the temperature anisotropies, but nucleosynthesis can easily be disrupted by departures from thermal equilibrium when $T \sim 0.1$ MeV \cite{Olive:1999ij}.   Given the  prefactor of $1/6$ on the $H_{\rm  reh}$ term, the effective difference between these two scales is very small.  Thus while one might take $T_{\rm  reh} \sim 0.1$ MeV as a physical lower bound on reheating, it is only slightly more restrictive to impose  $T_{\rm reh} \gsim  10$ MeV,  which also guarantees a thermalized neutrino sector.

\begin{figure}[tbp]
\includegraphics[width=\textwidth]{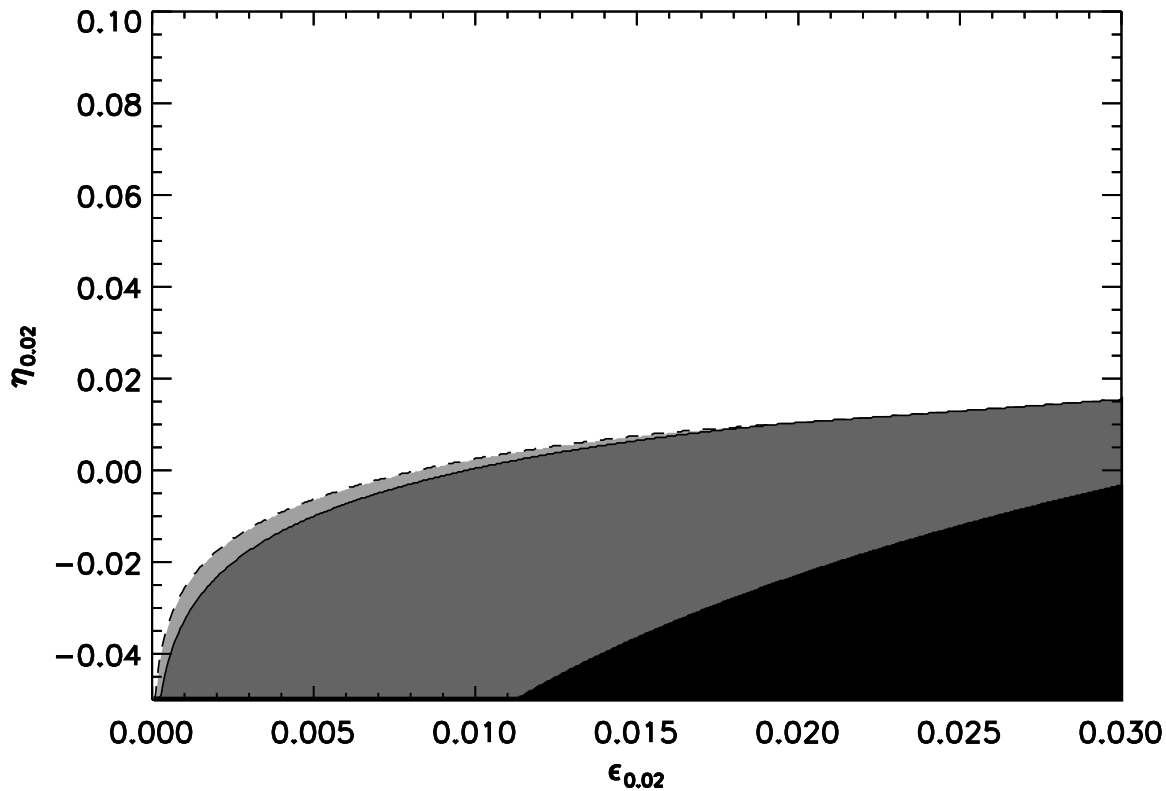}
\caption{The regions of the $\{ \epsilon,\eta\}$ plane excluded by requiring $N(k_\star)>15$ (black), $T_{\rm reh} > 10$ TeV (dark) and instantaneous thermalization (light). The other prior used in this work,  $T_{\rm reh} > 10$ MeV, looks very similar to the $T_{\rm reh} > 10$ TeV prior. For this figure only, the amplitude for the scalar spectrum at the fiducial scale has been set to the WMAP5 best fit. \label{fig:prior}}
\end{figure}

We use this information to eliminate combinations of slow roll parameters that cannot produce a sufficient amount of inflation, assuming that the same number of parameters used to fit the data provide a description of the inflaton dynamics till the end of inflation.  As the above analysis shows, this quantity depends on $H_k$, $H_{\rm end}$, and our assumptions about reheating.  We could also reject models that produce too much inflation -- but this assumes that inflation ends via the gradual ``erosion'' of slow roll, rather than via a hybrid transition -- a valley or a cliff, in the terminology of \cite{Adshead:2008vn} -- and we do not make this cut, instead treating equation~(\ref{lnconnect}) as a lower bound on $N$.  Physically, we are assuming that  the equation of state between the end of inflation and reheating is somewhere between $0$ and $1/3$. We are  thus ruling out a secondary period of accelerated expansion such as thermal inflation \cite{Lyth:1995ka} or post-inflationary cosmic string networks \cite{Burgess:2005sb} -- both of which would reduce $N$. Conversely, a ``kination'' dominated phase where $w\sim1$ would increase  $N$ beyond the lower bound we compute \cite{Chung:2007vz}. In a given model of inflation, $N$ is maximized when $H_{\rm reh} = H_{\rm end}$ and the universe thermalizes instantaneously after inflation.    Conversely, if thermalization happens slowly, we still need  $T_{\rm reh} \sim 10$~MeV.  Here we compute $H$ by assuming $\rho_{\rm reh} \sim T_{\rm reh}^4$. This is enough for our purposes, since the numerical factor which depends on the number of degrees of freedom appears logarithmically.   In Figure~\ref{fig:prior} we show the exclusion regions found for the $\{\epsilon,\eta\}$  parameter space with three versions of  the prior -- we will see later that this cut is visible in the two parameter $\{\epsilon,\eta\}$ chains.

In what follows, we explore the consequences of four different assumptions about the duration of inflation. The first is the bare prior that the inflation lasts long enough for modes which contribute to the CMB to leave the horizon -- without this we cannot self-consistently use the inflationary expressions for the power spectrum, and it amounts to demanding that $N>15$.\footnote{This number is somewhat arbitrary, but it cannot be made significantly lower while simultaneously ensuring that the perturbation spectrum is computed self-consistently. 
}  The next weakest constraint is to demand $T_{\rm reh} > 10$~MeV, which protects nucleosynthesis and guarantees  a thermalized neutrino population.  A stronger assumption is to require   $T_{\rm reh} > 10$~TeV, which ensures reheating occurs well above the electroweak scale. Finally, we consider ``instant'' reheating where $H_{\rm reh} = H_{\rm end}$.  Since $H$ decreases strictly with time,  the following sequence of inequalities must hold:
\begin{equation}
H_k  > H_{\rm end } \ge H_{\rm reh} ,
\end{equation}
where the first relationship is guaranteed by the dynamics of inflation. 
If we assign a  numerical value to $T_{\rm reh}$, we must enforce the second inequality explicitly.   Since we put a flat prior on $\epsilon$, our chains typically sample very few points with $\epsilon \lesssim 10^{-5}$.  The initial value of $\epsilon$ and the observed amplitude of the spectrum fix $H_k$,  so models with low $\epsilon$ also have a low inflationary scale. For the explicit values of $T_{\rm reh}$ we use, this constraint is always satisfied in practice. However, if one puts a  logarithmic prior on $\epsilon$, $H_k$ can be minute -- and any numerical bound on $T_{\rm reh}$ implicitly provides a lower bound on  $\log(\epsilon)$.  When $\epsilon$ is very small, it effectively decouples from the slow roll hierarchy \cite{Adshead:2008vn}, so  parameter regions found with a flat prior on $\log{\epsilon}$ would be determined by the prior and not the data. Consequently,  we do not pursue this question here.

\section{Parameter Sets and Monte Carlo Markov Chains}

We carry out MCMC studies of two sets of ``primordial'' parameters: $\{\epsilon, \eta, \log [10^{10} A_s] \}$ and $\{\epsilon, \eta, \xi, \log [10^{10} A_s] \}$ with flat priors  on each of these variables, and  the standard ``late-time'' cosmological parameters are also allowed to vary in the chains. In the language of \cite{Adshead:2008vn} these are High-$\epsilon$ 2-Parameter and High-$\epsilon$ 3-Parameter models, respectively.  The first parameterization is roughly analogous to the standard $\{ r, n_s, \log [10^{10} A_s]\}$ formulation; the second corresponds to the extended $\{ r, n_s, d n_s/d\ln k, \log [10^{10} A_s]\}$ set.    
We constrain these models with three data combinations: WMAP 5 year data (WMAP5), WMAP5 combined with the ACBAR 2008 release (WMAP5 + ACBAR) and WMAP5 combined with the SuperNova Legacy Survey (WMAP5 + SNLS).\footnote{We checked the combination WMAP5+ACBAR+SNLS but found that the ACBAR data did not significantly change results relative to WMAP5+SNLS and thus do not present those constraints.}  We do not marginalize over the amplitude of the Sunyaev-Zel'dovich (SZ) effect with a flat prior and an SZ template for a fixed set of cosmological parameters as was done with WMAP5 in \cite{Dunkley:2008ie}. In principle each vector of cosmological parameters visited by the chains has a unique SZ spectrum which should to be added to the primordial $C_\ell$ before being compared with the data, but computing  the SZ templates is not a trivial task. In practice, for all relevant parameter fits in \cite{Dunkley:2008ie}, the posterior of the SZ amplitude parameter essentially reproduces the prior, suggesting the SZ effect is not detected in WMAP5. Conversely, marginalization over the SZ parameter can potentially bias other parameters via volume effects, which  depend on the chosen width of the prior.  For these reasons we choose not to include an SZ parameter in our chains.   For consistency, we only consider ACBAR bandpowers at $\ell < 1800$, since higher multipoles are potentially SZ-contaminated. A summary of our parameter fits is given in Table~\ref{table:summary}.  

\begin{table}[!tb]{
 \begin{tabular}{||c|c|c||}
\hline
  Primordial Parameters & Dataset Combination & e-fold Prior   \\
\hline
$\{\epsilon, \eta, \log [10^{10} A_s] \}$ &  WMAP5  & $N_\star>15$ \\
$\{\epsilon, \eta, \xi, \log [10^{10} A_s] \}$ & WMAP5 + ACBAR  & $T_\mathrm{reh}> 10\ \mathrm{MeV}$    \\
 & WMAP5 + SNLS &  $T_\mathrm{reh}> 10\ \mathrm{TeV} $\\
 & & $H_\mathrm{end} = H_\mathrm{reh} $\\
\hline
\end{tabular}
 \caption{ MCMC parameter fits  described in this paper.   We run chains for all combinations of variables, dataset and e-fold prior.
}
\label{table:summary}
}
\end{table}

We use modified versions of the Boltzmann code CAMB \cite{Lewis:1999bs} and the public COSMOMC code \cite{Lewis:2002ah} for our MCMC parameter estimation. Each run uses eight chains. To assess the convergence of the chains, we apply two stringent requirements: a Gelman \& Rubin \cite{Gelman92} criterion on the least-converged eigenmode of the full parameter covariance, $R-1 <0.01$, and a conservative criterion on the convergence of confidence limits, stopping when the worst rms difference in the 95\% CL between chains for all parameters is 10\% of the standard deviation. 

\begin{figure}[!thbp]
\includegraphics[width=\textwidth]{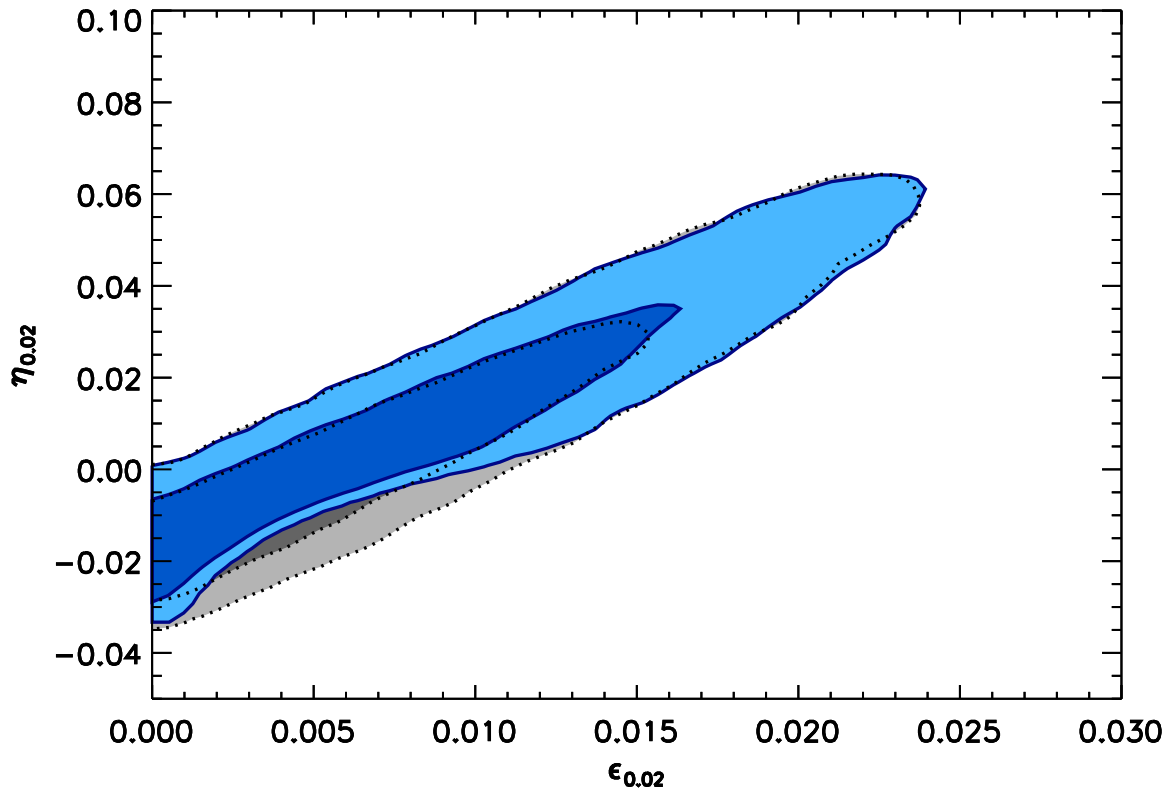}
\caption{The joint 68\% (dark) and 95\% (light) confidence levels obtained on the first two HSR parameters $\epsilon$ and $\eta$ at the fiducial scale $k_\star=0.02$ Mpc$^{-1}$ from the WMAP 5 year data, assuming $\xi$ and all higher order slow roll parameters are zero. The grey (dotted) contours correspond to a ``minimal'' e-fold prior, $N(k_\star)>15$. The blue (solid) contours are obtained by assuming instantaneous reheating.  \label{fig:fr2}}
\end{figure} 

\begin{figure}[!thbp] 
\includegraphics[width=5in]{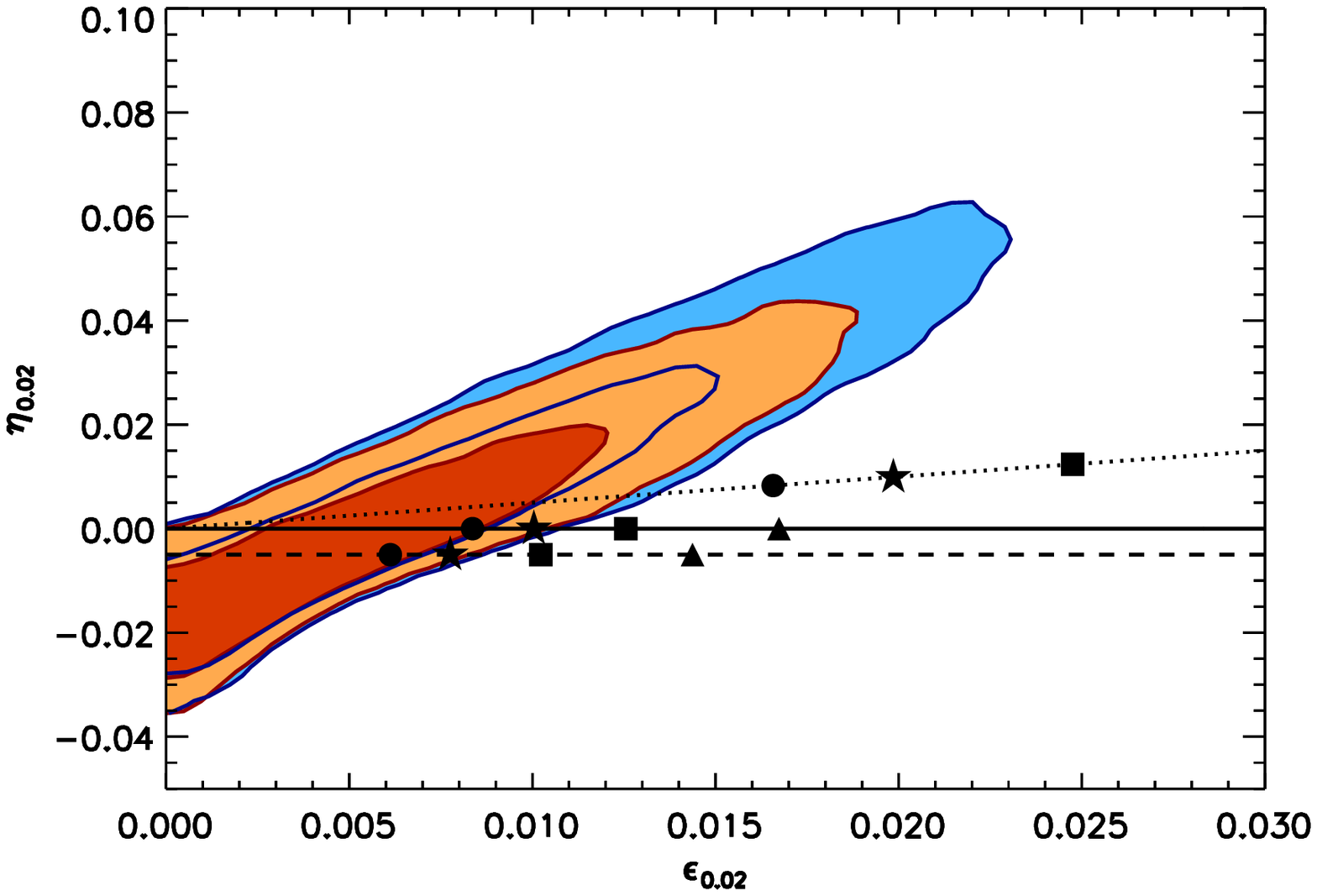}
\includegraphics[width=5in]{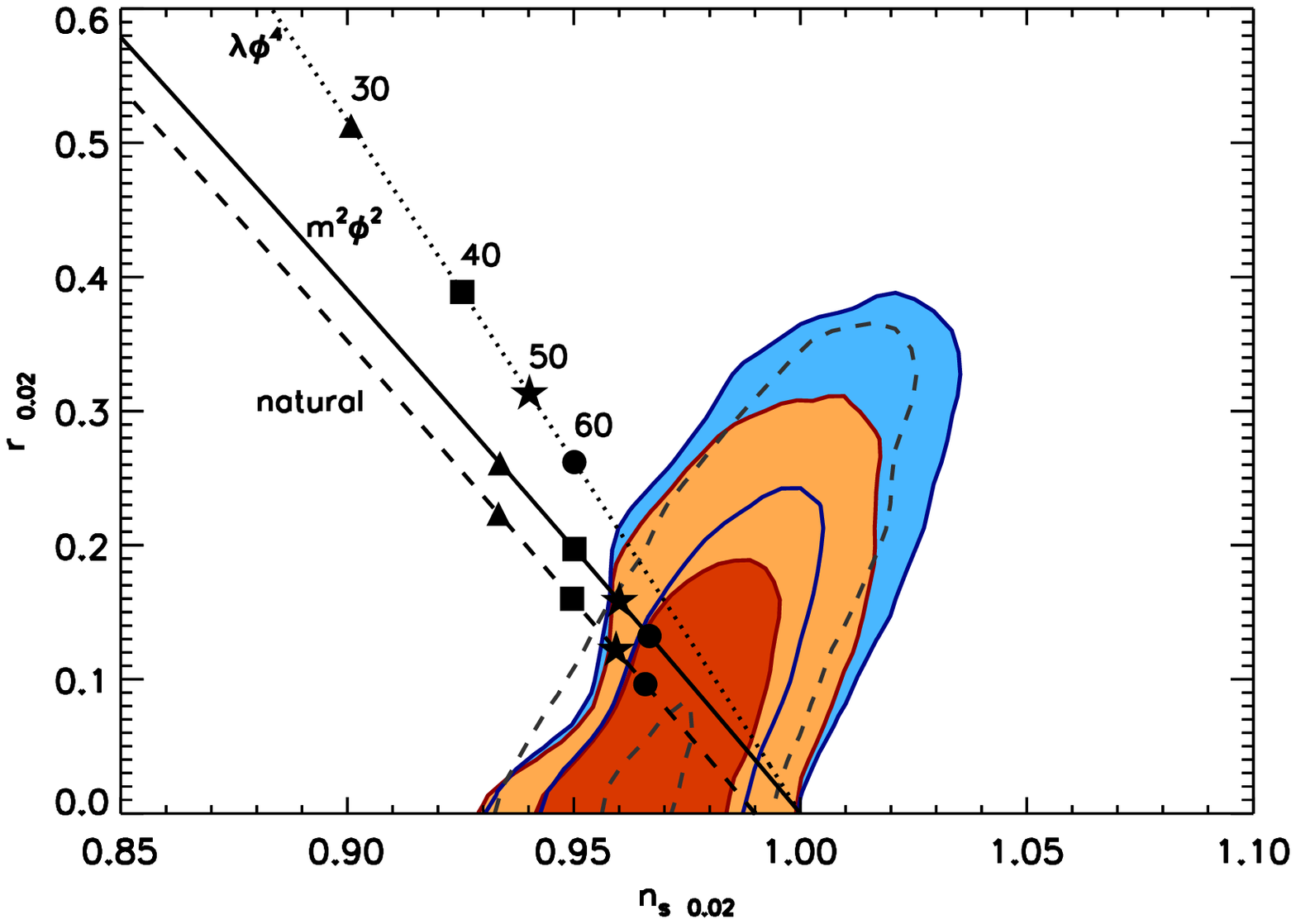}
\caption{The joint 68\% (inner) and 95\% (outer) bounds on  the slow roll variables (top) and power law spectral parameters (bottom),  with $k_\star = 0.02\ \mathrm{ Mpc}^{-1}$, for a High-$\epsilon$ 2-parameter fit.  WMAP5 constraints are blue, and WMAP5+SNLS  constraints are red. Solid contours come from Slow Roll Reconstruction with  prior $T_\mathrm{reh} > 10$ TeV. The running of the scalar index in these models is second order in slow roll and hence very small. The dashed contours show results from WMAP5 chains where the spectrum was specified via $n_s$ and $r$. We superimpose the ``trajectories'' for three generic  slow roll models, $\lambda \phi^4$ (dotted), $m^2 \phi^2$ (solid), and a representative natural inflation model (dashed), along with the position at different values of $N=30$ (triangle), $N=40$ (square), $N=50$ (star), $N=60$ (circle). \label{fig:fr2zoo}}
\end{figure}

Our constraints on the inflationary parameters for different data combinations and various reheating priors are presented in Tables \ref{table:hsr_constraints_fr2} and \ref{table:hsr_constraints_fr3}. The WMAP5+SNLS data combination is significantly more constraining than WMAP5+ACBAR. We refer the reader to these tables for quantitative details, and now highlight the key points. 
The direct effect of the reheating prior on the $\{\epsilon, \eta, \log [10^{10} A_s] \}$ parameterization is a ``bite'' into the constraints,  as displayed in Figure \ref{fig:prior}. We illustrate the impact of the choice of reheating prior on the parameter constraints derived for WMAP5 in Figure \ref{fig:fr2}\footnote{The keen-eyed reader may notice that the cut in Figure \ref{fig:prior} is slightly higher than what appears in the constraints in Figure \ref{fig:fr2}. This is an unavoidable artifact of the binning used to calculate 2D contours from MCMC using the public GetDist statistics package. The MCMC itself reproduces the sharp cut of the prior, but the binning aliases this curve to slightly lower values.}. In Figure \ref{fig:fr2zoo}, we show the constraints in the $\{\epsilon, \eta\}$ plane for a representative (10 TeV) reheating prior, for both the WMAP5 and WMAP5+SNLS data combinations. We transform these constraints into the empirical $\{r, n_s\}$ parameters by using the approximate second order slow roll formulae, and compare these to  constraints obtained  directly from an MCMC analysis where  $\{r, n_s\}$ were varied in the chains. The difference follows from the choice of e-fold prior: if the prior eliminates regions which would otherwise have a high likelihood, the resulting confidence levels will stretch further into the tails of the likelihood distribution, where the likelihood is somewhat lower. This can lead to a broadening of the marginalized constraints if the slope of the likelihood surface is relatively shallow in the tails, and this effect can be seen clearly in the 68\% contour in Figure~\ref{fig:fr2}.  
We also show trajectories as a function of the number of e-folds for three common slow roll models. The $\lambda \phi^4$ ``toy-model'' is firmly excluded, while $m^2 \phi^2$ and natural inflation models are still allowed at the 95\% CL level for $N>50$ -- which accords with the number of e-folds we would expect for these models.  We find that our WMAP5+SNLS analysis disfavors a blue spectral index even when $r$ (or $\epsilon$, in our case) is included in the chains, which accords with the  WMAP Team conclusions for  the data combination of WMAP5, a compilation of supernovae data and baryon acoustic oscillations data \cite{Komatsu:2008hk}. 

\begin{figure}[!thbp] 
\includegraphics[width=5in]{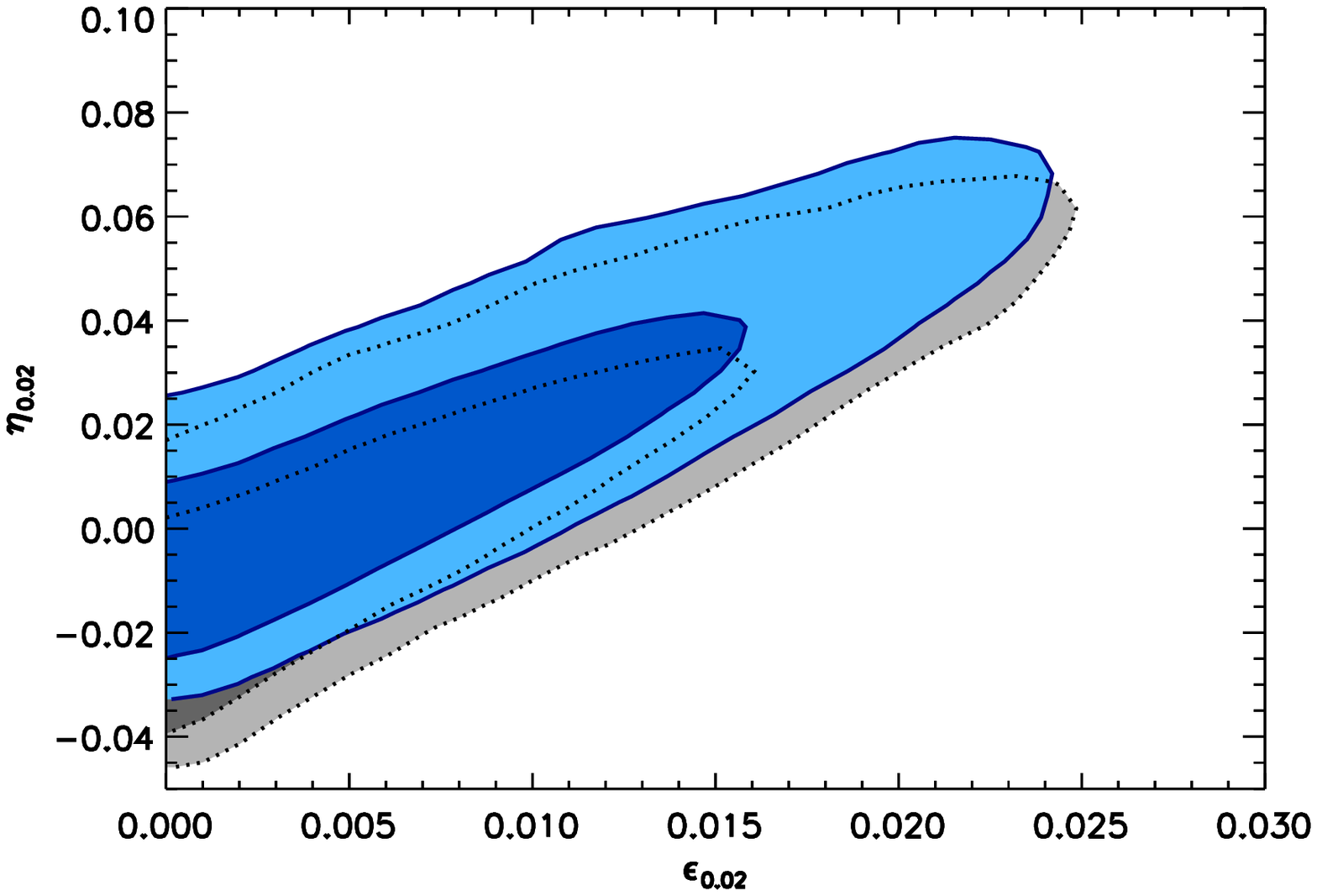}
\includegraphics[width=5in]{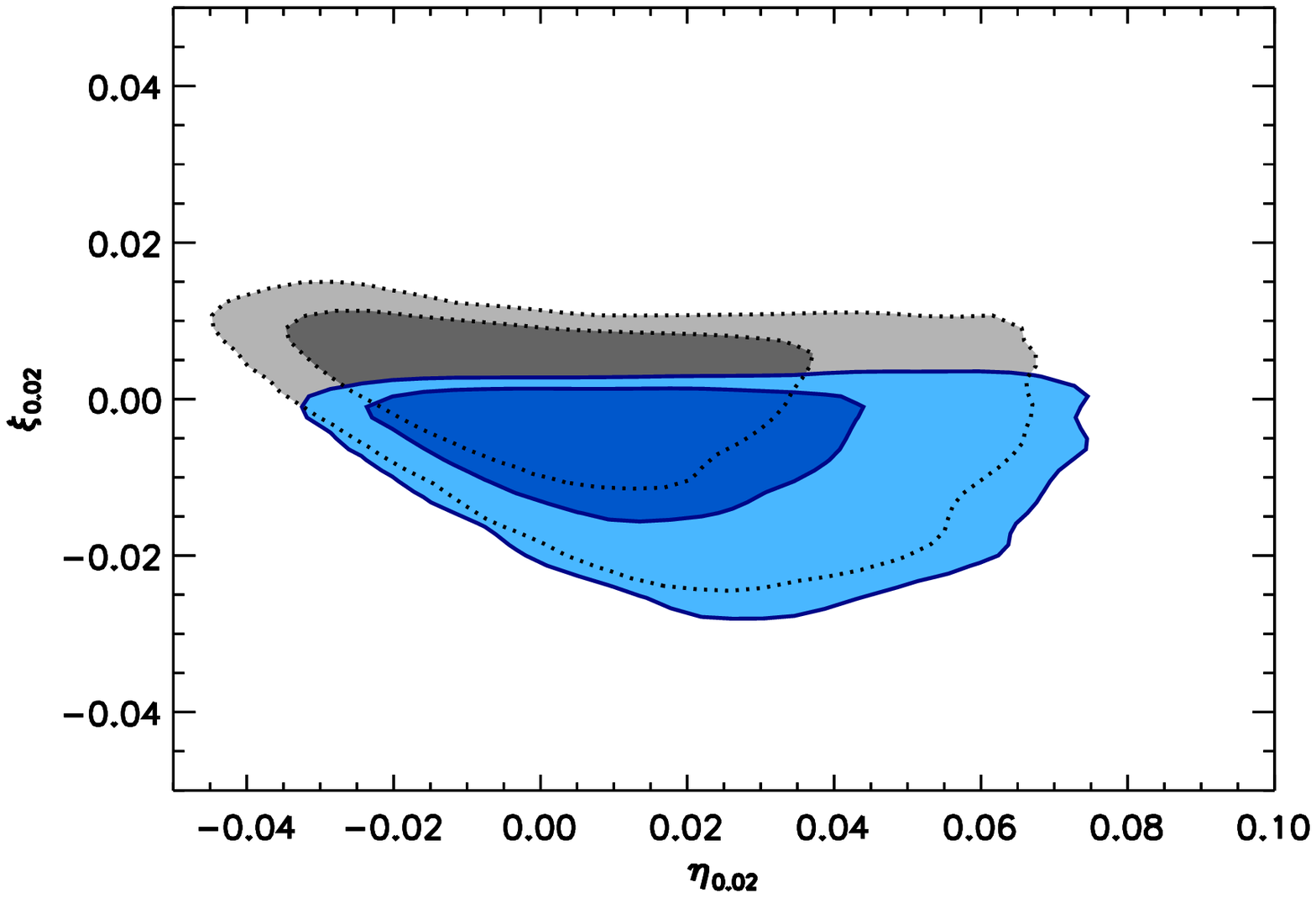}
\caption{The joint 68\% (dark) and 95\% (light) confidence levels obtained on the first three HSR parameters $\{\epsilon, \eta, \xi\}$ at the fiducial scale $k_\star=0.02$ Mpc$^{-1}$ from the WMAP 5 year data, assuming all higher order slow roll parameters are zero. The grey (dotted) contours correspond to a ``minimal'' e-fold prior, $N(k_\star)>15$. The blue (solid) contours are obtained by assuming instantaneous reheating.  \label{fig:fr3priors}}
\end{figure} 

\begin{figure}[!thbp] 
\includegraphics[width=5in]{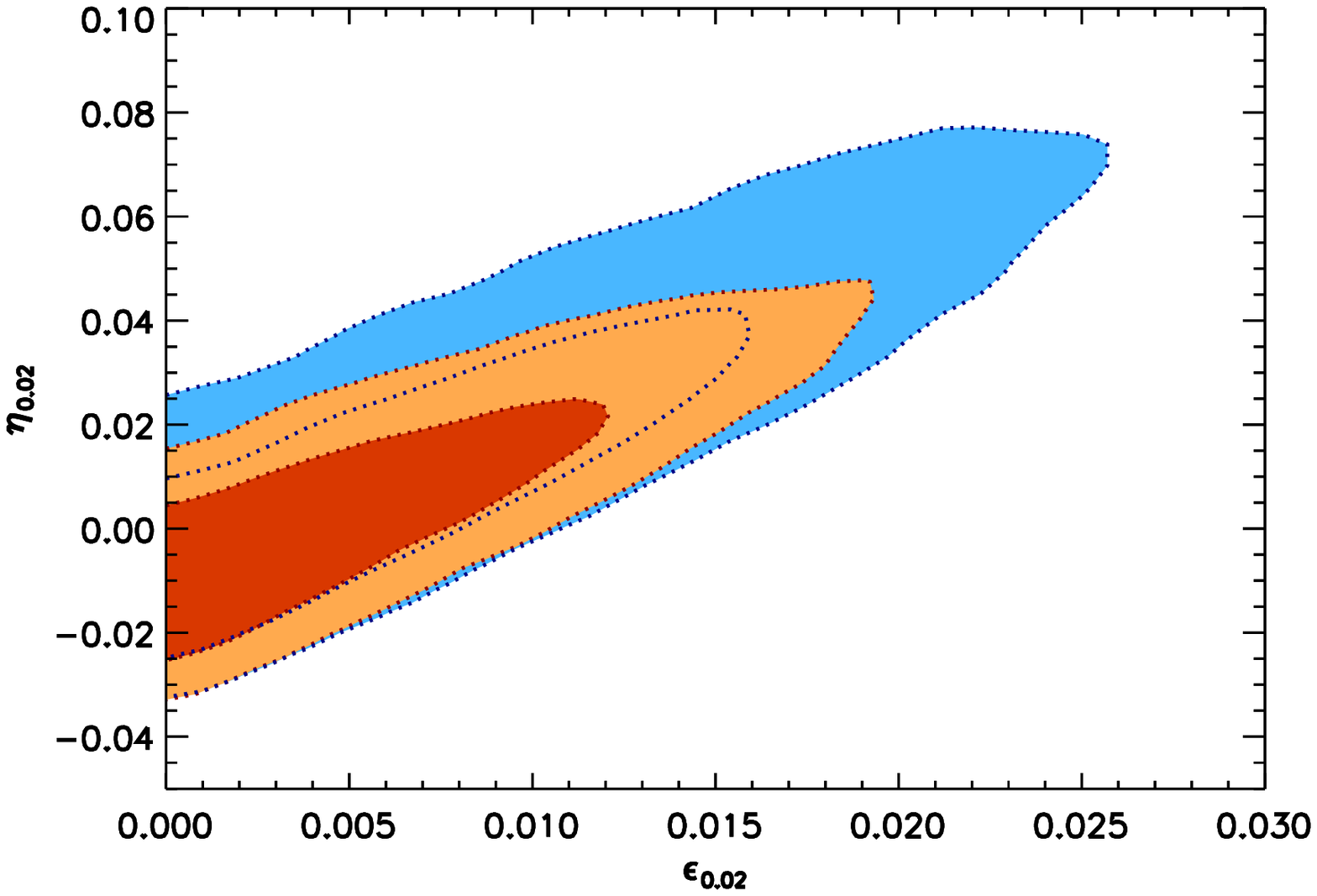}
\includegraphics[width=5in]{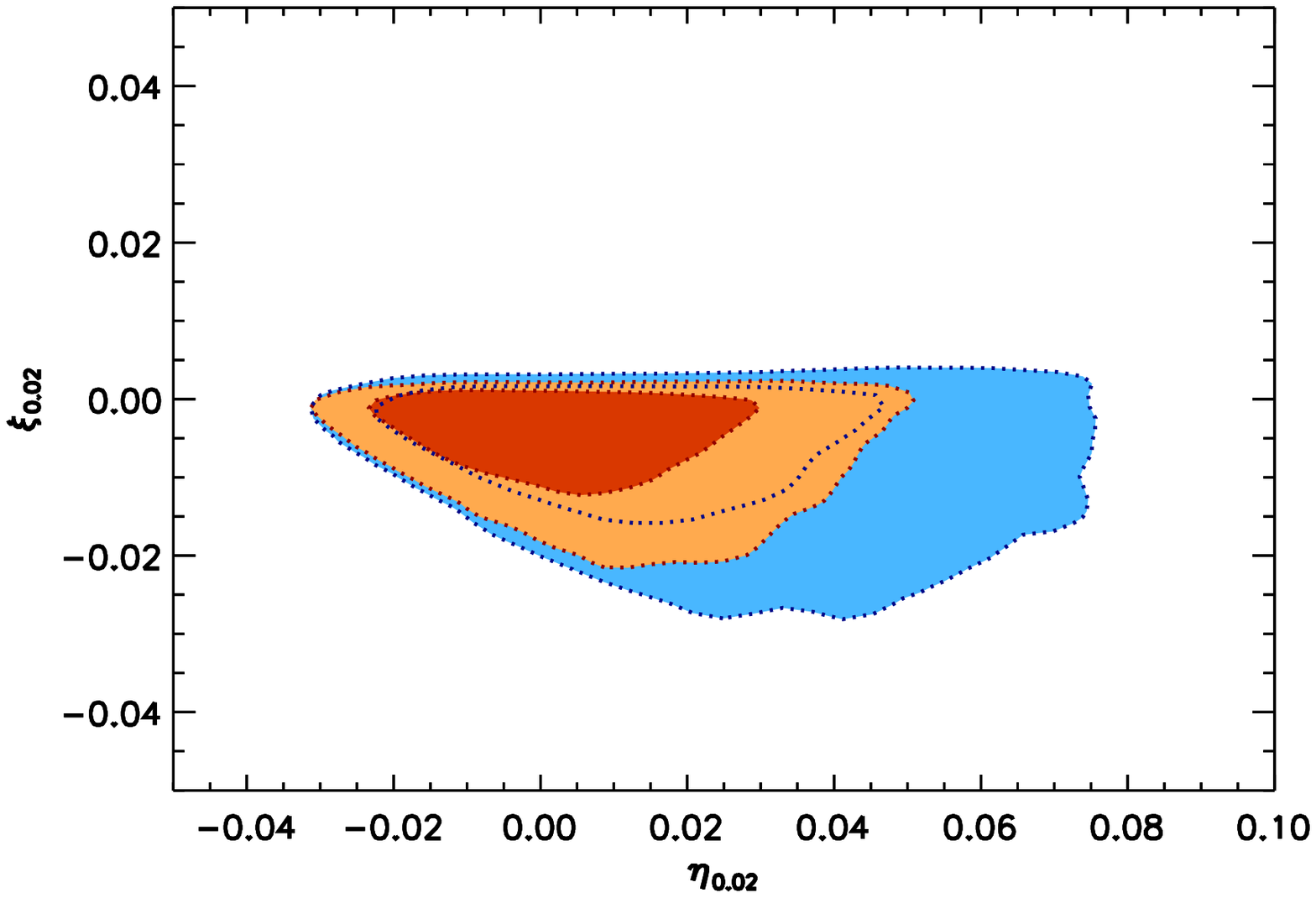}
\caption{The joint 68\% (inner) and 95\% (outer) bounds on the first three HSR parameters $\{\epsilon, \eta, \xi\}$ at the fiducial scale $k_\star = 0.02\ \mathrm{ Mpc}^{-1}$.  The blue constraints are derived from WMAP 5 year data alone, and the red constraints from the WMAP5+SNLS data combination, with  $T_\mathrm{reh} > 10$ TeV. 
\label{fig:fr3a}}
\end{figure}

\begin{figure}[!thbp] 
\includegraphics[width=5in]{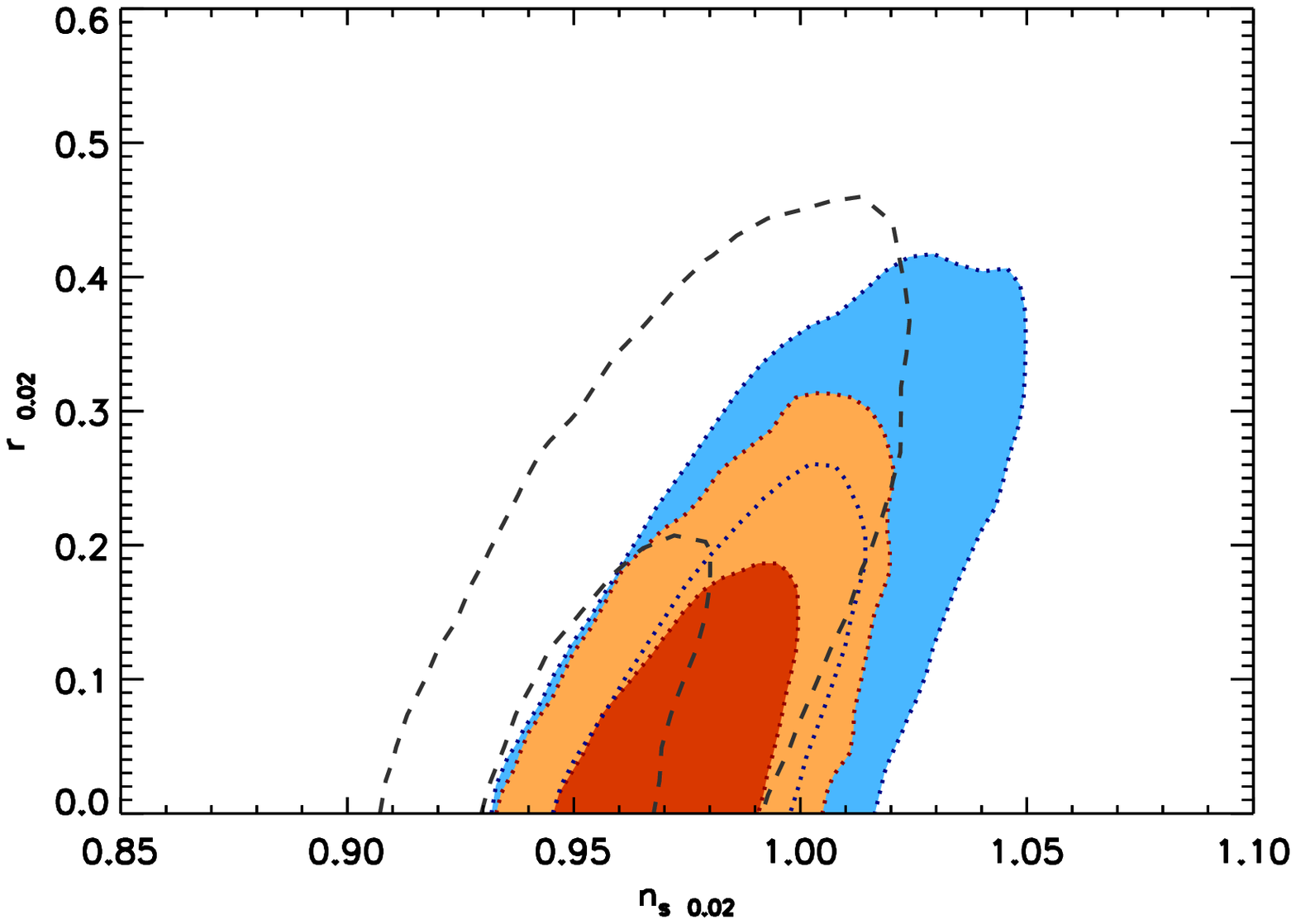}
\includegraphics[width=5in]{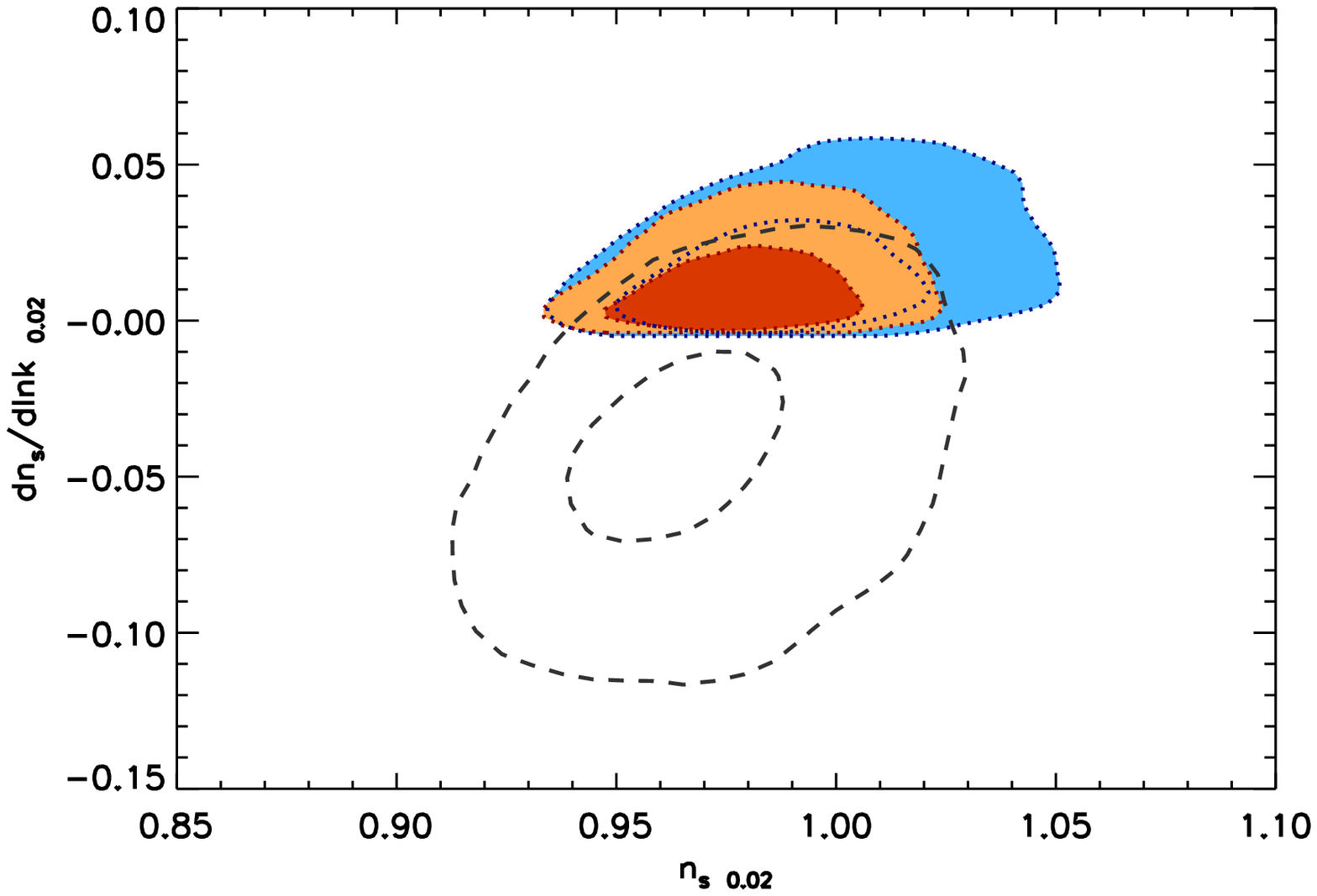}
\caption{The joint 68\% (inner) and 95\% (outer) bounds on the power law spectral parameters at the fiducial scale $k_\star = 0.02\ \mathrm{ Mpc}^{-1}$, obtained by transforming the constraints shown in Fig.~\ref{fig:fr3a} into this parameter space.  The blue constraints are derived from WMAP 5 year data alone, and the red constraints from the WMAP5+SNLS data combination. The dotted contours come from Slow Roll Reconstruction applying an e-fold prior assuming $T_\mathrm{reh} > 10$ TeV. For comparison, the dashed contours show an analysis using the empirical power-law prescription in terms of $n_s$, $r$, and $d n_s/d\ln k$ at a pivot scale of $0.02$ Mpc$^{-1}$, using WMAP5 data.  \label{fig:fr3b}}
\end{figure}

Figure \ref{fig:fr3priors} shows the difference between a minimal e-fold prior and the strongest (instantaneous) reheating prior for the $\{\epsilon, \eta, \xi, \log [10^{10} A_s] \}$ parameterization. The e-fold priors tend to reject models with  positive $\xi$, or negative running.  Even a minimal ($N_\star>15$) prior puts tight restrictions on the allowed upper limit of $\xi$.  Adding a low ($> 10$ MeV) reheating prior tightens this effect substantially, but adding a stronger reheating prior does not affect this limit further, implying that inflation ends so quickly for these models that the weakest reheating prior is sufficient to cut them out. We will return to this topic in Section \ref{sec:bh}.   Figure \ref{fig:fr3a} shows constraints on $\{\epsilon,\eta,\xi\}$ for  a  representative (10 TeV) reheating prior  for  WMAP5 and WMAP5+SNLS.    Figure \ref{fig:fr3b} shows  constraints for the same chains, after being post-processed (maintaining flat priors on the HSR parameters) to express them in terms of empirical power law parameters, using  second-order slow roll formulae.  We compare these results to a standard analysis using the power law parameters $\{n_s, r, dn_s/d\ln k\}$ in the chains. We see that the e-fold prior excludes most of parameter space allowed by naively fitting $dn_s/d\ln k$ to the data. As is now well-known, a large negative running (corresponding to large positive $\xi$) does not provide sufficient inflation. Marginalizing over this degeneracy with $dn_s/d\ln k$ shifts the constraints on $n_s$ and $r$ from the empirical analysis to redder spectral indices compared to Slow Roll Reconstruction with a reheating prior. The difference between the empirical constraints shown in Figure \ref{fig:fr3b} and those from the WMAP Team's analysis in \cite{Dunkley:2008ie, Komatsu:2008hk} is due to the fact that they used a pivot scale of $0.002$ Mpc$^{-1}$ (compared with our fiducial scale $0.02$ Mpc$^{-1}$), which exacerbates this degeneracy and hence shifts the marginalized constraints through a volume effect. 

The exclusion of large positive $\xi$ along this degeneracy by the e-fold prior again causes our MCMC to explore parts of likelihood space which were not within the marginalized constraints before this prior was applied, and allows a significant {\sl negative} $\xi$. We will explore the consequences of the lower bound on $\xi$ in the next section.

\section{Primordial Black Holes and the Risk of Eternal Inflation} \label{sec:bh}

In contrast to the constraints found with the WMAP3 dataset \cite{Peiris:2006ug,Easther:2006tv}, the WMAP5 results \cite{Komatsu:2008hk} show less evidence for a strong running of the scalar spectrum.   While second order terms in slow roll contribute to the running, these cannot be resolved with current astrophysical data \cite{Adshead:2008vn}. Therefore the running is dominated by the third slow roll parameter $\xi$, and in this limit, 
\begin{equation}
\alpha \equiv \frac{d n_s}{d\ln k } \approx -2 \xi \, .
\end{equation}
Large, positive values of $\xi$ lead to a very short period of inflation, and are thus excluded by an e-fold prior \cite{Easther:2006tv} and, as with WMAP3, the upper limit on $\xi$ derived from WMAP5 is driven solely by the e-fold prior. The lower limit on $\xi$ from WMAP5 is actually more negative than the analogous result from WMAP3 -- so the maximal permitted running is thus more positive.   The difference is primarily due to a better treatment of the beam and point sources in the latest release, rather than the extra two years of data.  This situation illustrates the importance of systematic effects,  and  reminds us of the need for caution when interpreting cosmological constraints.  

Interestingly,  significantly negative values of $\xi$ can also be excluded on physical grounds, again under the assumption that the first three HSR parameters give a description of the inflaton dynamics till the end of inflation.  With a  substantial negative  $\xi$, we find a class of solutions with $\epsilon \rightarrow 0$, and the field rolls towards a minimum with a substantial vacuum energy.  Given that  $\epsilon \rightarrow 0$ while $H$ remains finite, the amplitude of the perturbation spectrum diverges.    As $P_{\mathcal{R}}$ approaches unity (at small scales) primordial black holes can form in the post-inflationary universe \cite{Hawking:1971ei,Carr:1974nx,GarciaBellido:1996qt,Green:1997sz,Yokoyama:1999xi,Leach:2000ea,Chongchitnan:2006wx,Zaballa:2006kh,Kohri:2007qn}.  Their mass is limited by the energy contained within the horizon volume, which increases as the universe expands. A black hole formed at a  temperature greater than $\sim 10^8$ GeV decays before the present epoch.  Consequently, if {\em stable\/} black holes form after inflation with a high (near-GUT) reheat temperature, they are generated by modes which reenter the horizon with a substantial amplitude {\em after\/} the temperature has fallen below this critical value.  In particular,  \cite{Leach:2000ea,Chongchitnan:2006wx,Kohri:2007qn} consider primordial black hole formation after slow roll inflation, and the impact of a running scalar index driven by a large negative $\xi$ term.  Interestingly, \cite{Chongchitnan:2006wx} uses a version of Monte Carlo reconstruction  \cite{Kinney:2002qn,Easther:2002rw} to look for potentials whose perturbations have spectral indices consistent with CMB data while also producing large numbers of small black holes. The only potentials they find which fit both criteria have sharp ``features'', and thus do not fit naturally within the slow roll hierarchy. The authors of  \cite{Kohri:2007qn} note that this conclusion is at odds with \cite{Leach:2000ea}, and attribute the discrepancy to ``the use of flow equations and the hierarchy''\footnote{By ``the hierarchy'', which in standard usage means the infinite hierarchy of differential equations describing the evolution of the HSR parameters, the authors of  \cite{Kohri:2007qn} instead mean progressively tighter restrictions on the initial conditions for the HSR parameters as given by their equation (55). However,  \cite{Chongchitnan:2006wx} only apply these restrictions to the {\em priors} on their initial conditions. Individual draws from their priors are manifestly able to generate dynamics which violate these restrictions. Thus the objections of \cite{Kohri:2007qn} do not apply.} by \cite{Chongchitnan:2006wx}.  However, as we saw in Figures \ref{fig:fr3a} and \ref{fig:fr3b}, our chains -- which also use the flow equations -- have no difficulty finding regions of parameter space which lead to the overproduction of primordial black holes. 

Instead, the discrepancy is attributable to the way that ``Monte Carlo reconstruction'' \cite{Kinney:2002qn,Easther:2002rw} handles models in which inflation can continue indefinitely.  Namely, one assumes that if there is a late time attractor, the inflationary trajectory {\em reaches\/} this attractor, whereupon inflation ends via a hybrid transition.  As noted in \cite{Easther:2002rw}, this is a simplifying assumption. It ensures that the spectrum has a blue spectral index at CMB scales, since (in slow roll terminology) $V'$ is vanishingly small and $V''$ is negative, and these models are excluded by the data. Thus, one cannot find sets of slow roll parameters which overproduce primordial black holes while also matching the CMB power spectrum with this ansatz.  Conversely, Slow Roll Reconstruction works with the values of the slow roll parameters at CMB scales, and effectively marginalizes over the mechanism by which inflation ends. We thus recover models which have not quite reached the attractor solution $\epsilon \rightarrow 0$ at CMB scales, so their spectral index  is red with a large positive running. By the end of inflation, the spectra generated by these models are apparently large enough to source the production of primordial black holes, and this class of model is missed by the algorithm used by \cite{Chongchitnan:2006wx}. In other words, they make an arbitrary and non-unique choice for the piece of the trajectory which corresponds to the CMB scales which {\em a priori} rejects the models we have found which both match constraints at CMB scales and lead to primordial black hole overproduction by the end of inflation. The running mass model is an explicit inflationary scenario with these dynamics   \cite{Stewart:1996ey,Stewart:1997wg}, invoking softly-broken global supersymmetry during inflation. This model has been compared to observations in a number of works \cite{Covi:1998jp, Covi:1998mb, Leach:2000ea, Covi:2002th,Covi:2004tp}. 

\begin{figure}[!th] 
\includegraphics[scale=0.45]{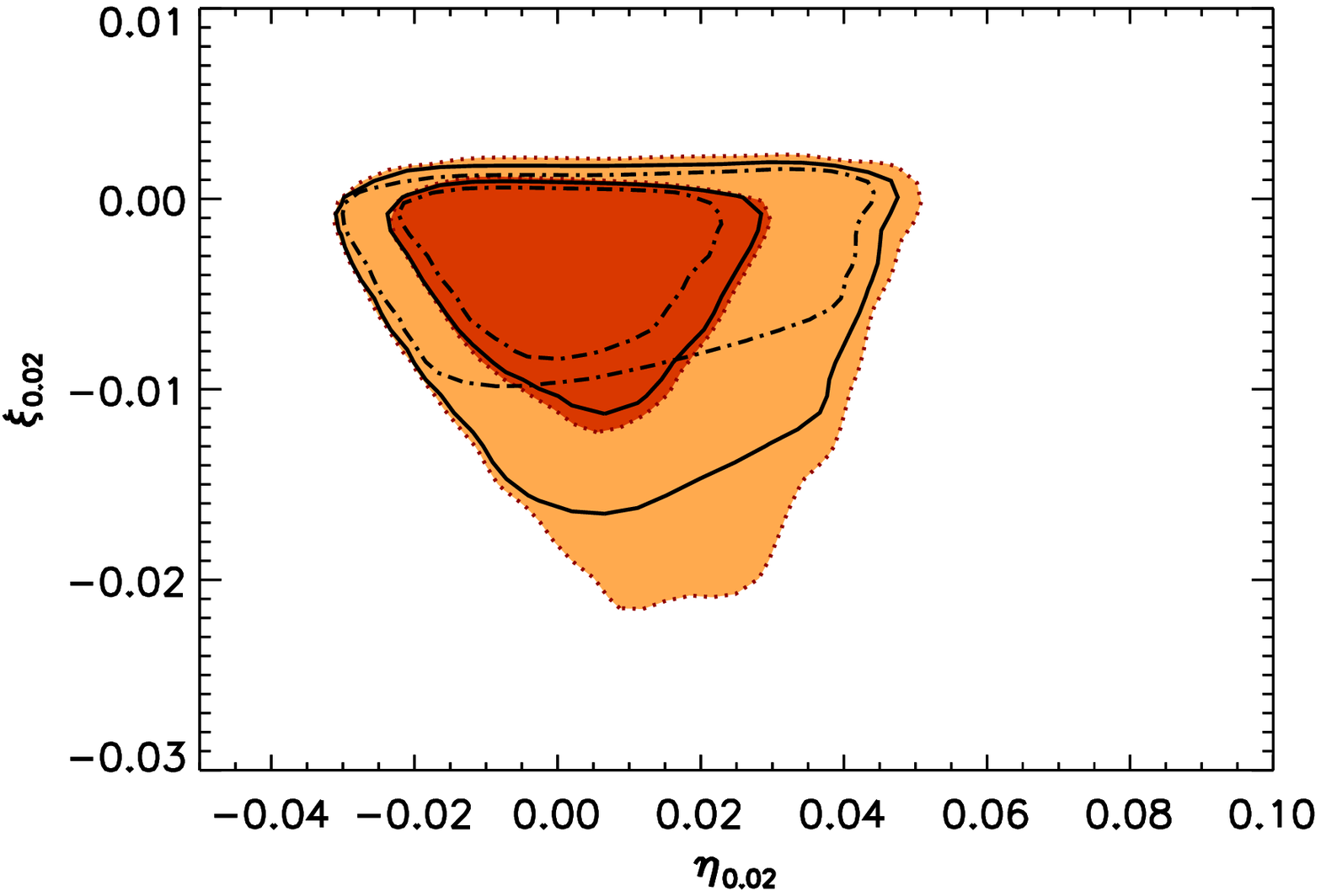}
\includegraphics[scale=0.45]{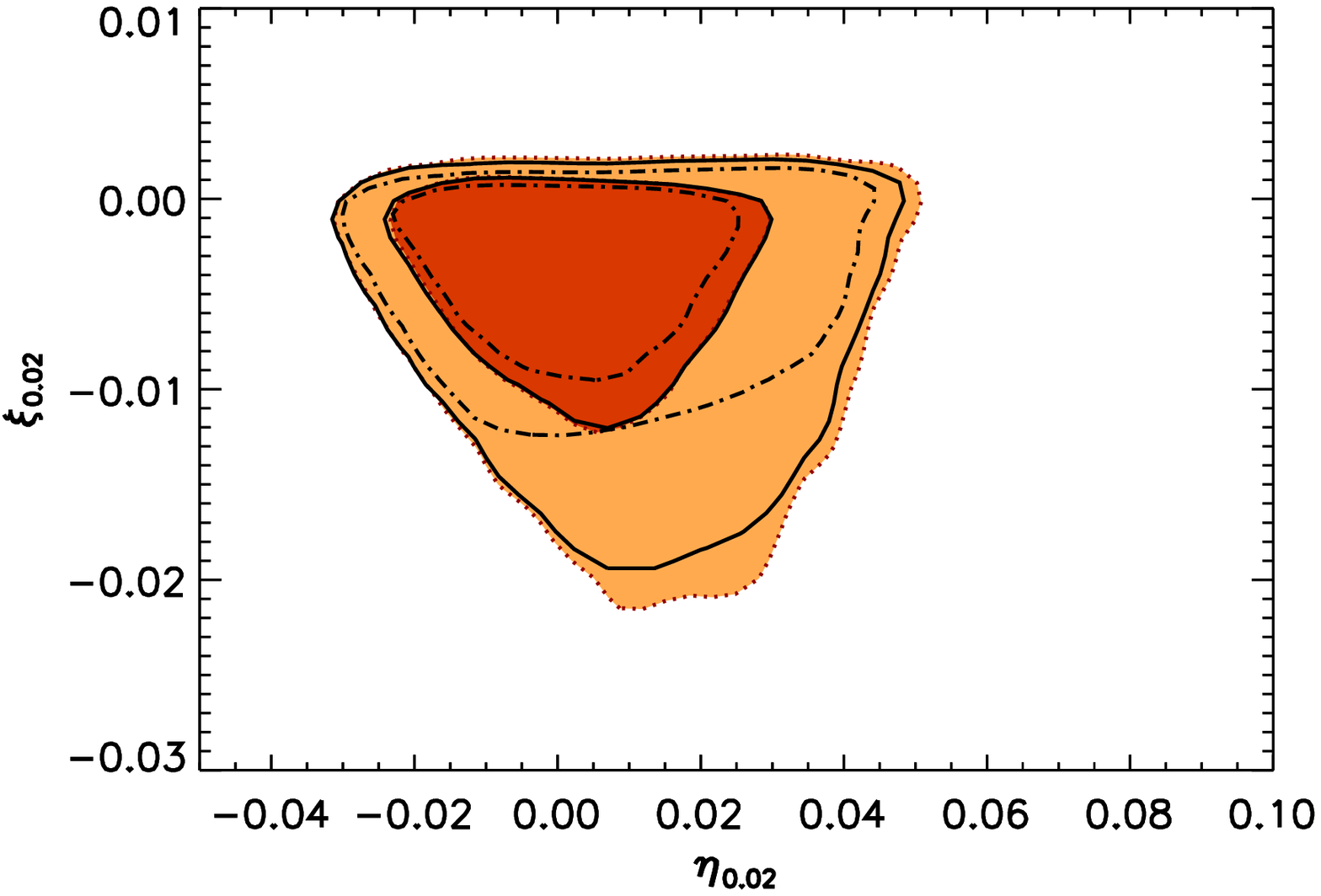} \\
\includegraphics[scale=0.45]{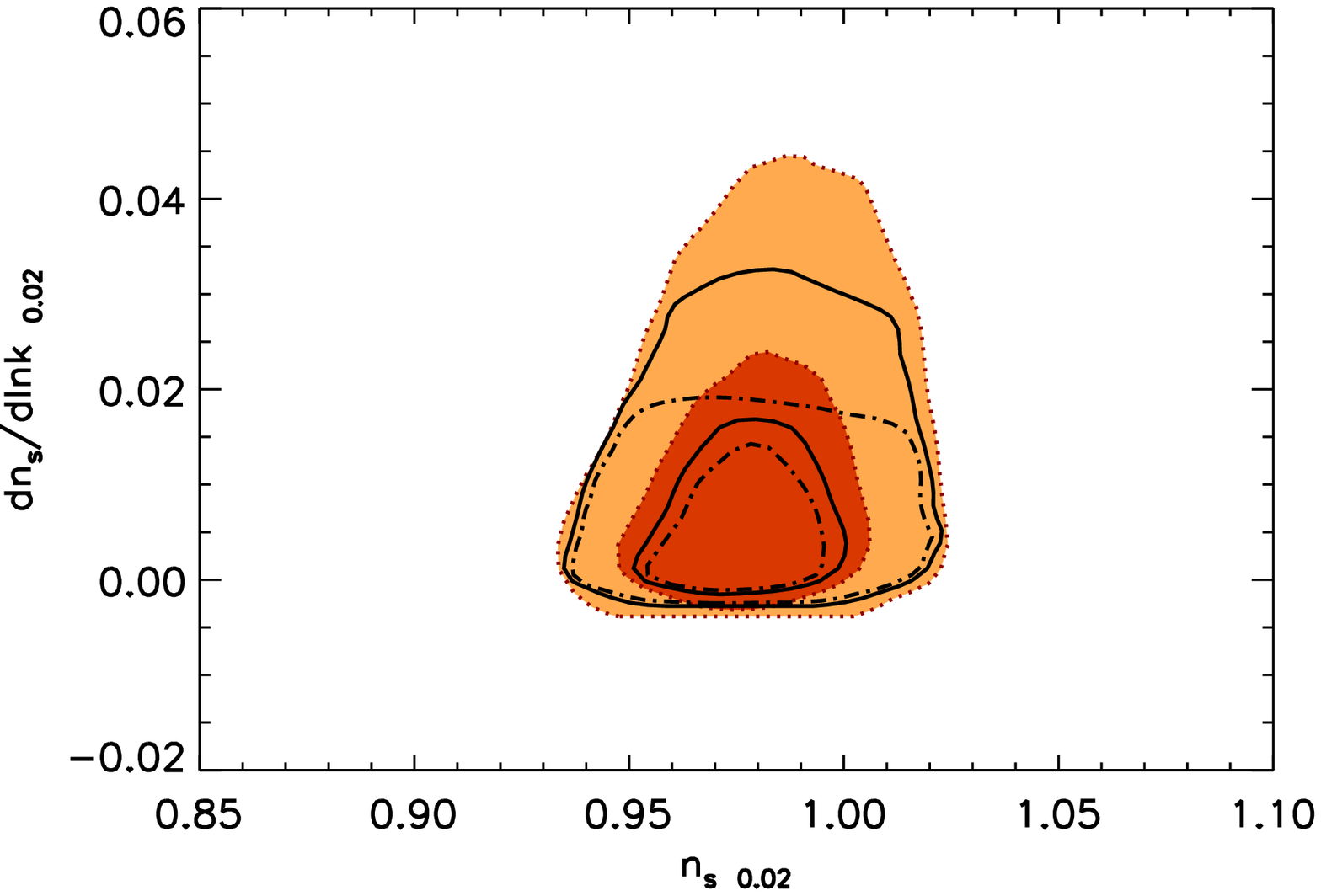} 
\includegraphics[scale=0.45]{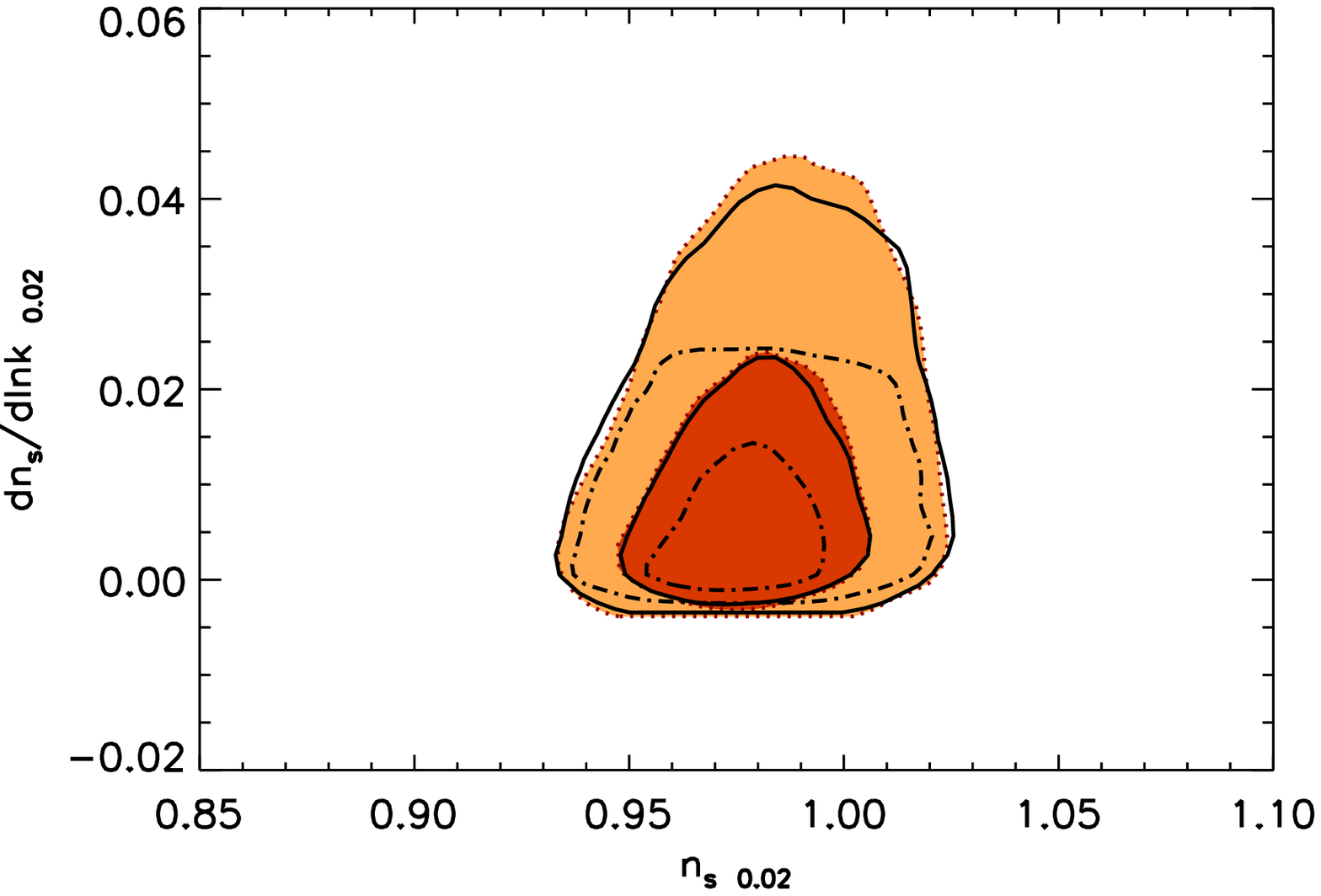}
\caption{The joint 68\% (inner) and 95\% (outer) bounds at the fiducial scale $k_\star = 0.02\ \mathrm{ Mpc}^{-1}$.  The red constraints come from the WMAP5+SNLS data combination using Slow Roll Reconstruction, applying an e-fold prior assuming $T_\mathrm{reh} > 10$ TeV. This e-fold prior eliminates models with large {\sl positive} $\xi$ and has no effect on models with large {\sl negative} $\xi$, which inflate forever in the absence of a hybrid transition to end inflation. The dark contours come from post-processing the same chains to exclude models which lead to primordial black hole overproduction (see text for discussion): solid black contours assume that $T_\mathrm{reh} = 10$ MeV, and the dot-dash contours assume instantaneous reheating. The left and right hand columns are based on requiring that the power spectrum amplitude at the end of inflation satisfy $P_\mathcal{R} < 10^{-3}$ and $P_\mathcal{R} < 10^{-2}$ respectively. We thus see that the black hole constraint depends strongly on the duration of inflation and the precise value of the amplitude at which excessive black hole production begins. \label{fig:fr3bh}}
\end{figure} 

In what follows, we examine the lower limit on $\xi$ that results from imposing an additional prior based on primordial black hole overproduction. The limit on the primordial amplitude at the end of inflation is somewhat uncertain -- a variety of arguments \cite{Carr:1994ar, Kim:1996hr, Green:1997sz, Green:1999yh, Lemoine:2000sq, Kohri:2007qn} suggest that $P_\mathcal{R} < 10^{-3}$--$10^{-2}$. Further, as we have seen, in the absence of a concrete mechanism to end inflation, there is a large uncertainty in the identification of where inflation ends in these models. As discussed previously (for example, by Leach and collaborators \cite{Leach:2000ea}), this uncertainty has a significant effect on the constraint as well. To take into account these uncertainties, we post-process our chains as follows.  We identify the end of inflation $N_\mathrm{end}$ using equation (\ref{lnconnect}) assuming a minimal $T_\mathrm{reh} = 10$ MeV and a maximal requirement of instantaneous reheating. Further, at $N_\mathrm{end}$, we apply two amplitude cuts: $P_\mathcal{R} < 10^{-3}$ and $P_\mathcal{R} <10^{-2}$. As pointed out in \cite{Leach:2000ea}, the Stewart-Lyth formulae are accurate enough to compute the power spectrum at the end of inflation in these models, provided that the background solution is computed exactly, as we have done. We show our results in Figure \ref{fig:fr3bh}.   We see that the lower limit on $\xi$ is strongly dependent on the precise form of the prior. For all but the weakest combinations of priors, a significant part of the observationally--allowed parameter space is excluded by primordial blackhole overproduction.

\begin{figure}[!thbp] 
\includegraphics[width=5in]{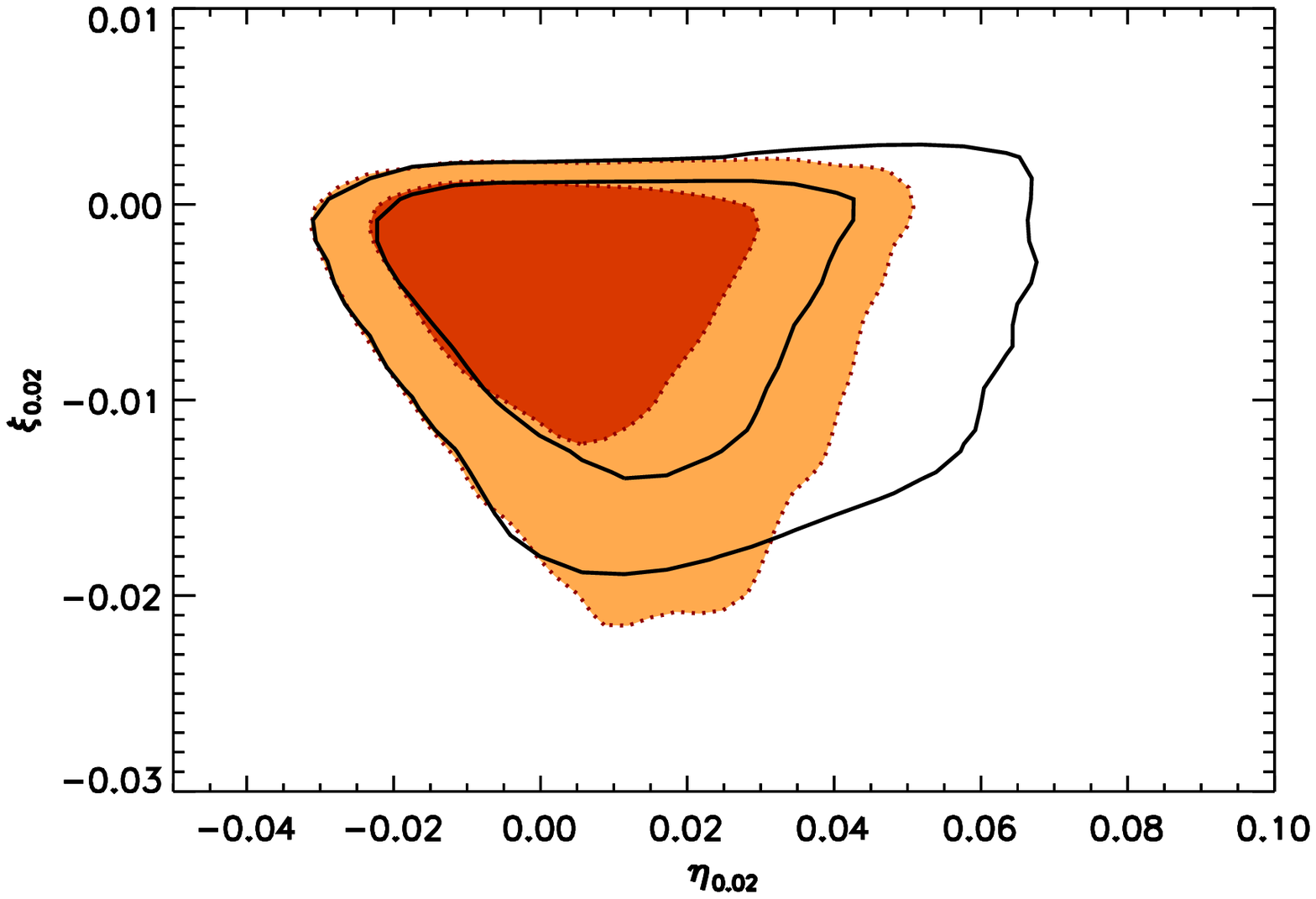}
\includegraphics[width=5in]{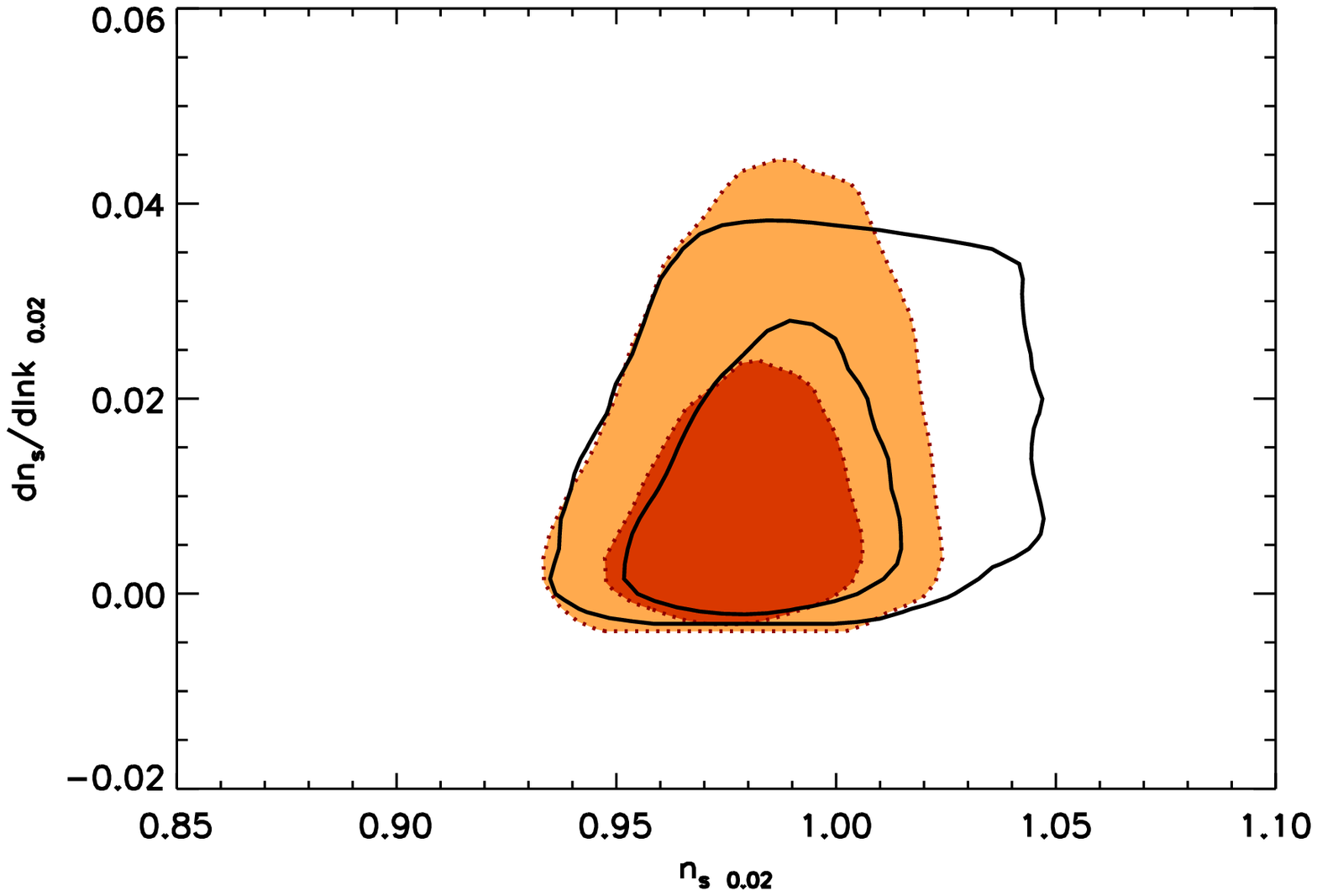}
\caption{The joint 68\% (inner) and 95\% (outer) bounds at the fiducial scale $k_\star = 0.02\ \mathrm{ Mpc}^{-1}$.  The red constraints come from the WMAP5+SNLS data combination using Slow Roll Reconstruction, applying an e-fold prior assuming $T_\mathrm{reh} > 10$ TeV. The solid black contours come from post-processing the equivalent WMAP5-only chains to impose an additional prior based on avoiding eternal inflation (see text for discussion).  \label{fig:fr3ei}}
\end{figure} 

Even if we ignore the previous constraint (though it should be clear that the following case is subset of the former), eternal inflation occurs in models where the quantum fluctuations (which are proportional to $H$) dominate the classical rolling, and this can be shown to be equivalent to the density perturbations being of order unity \cite{Linde:1986fd, Kohri:2007qn}. In this case we could not trust the usual classical inflaton equations of motion, which are embodied in the slow-roll hierarchy, and inflation cannot end coherently. This again rules out large, negative values of $\xi$ in three parameter models. The eternal inflation prior is implemented as follows.  First, we note that in the models where there is a danger of eternal inflation, $\epsilon \rightarrow 0$ while $H(\phi)$ asymptotes to a constant, $H_\mathrm{asymp}$. As we approach this limit, the primordial power spectrum $H^2/\epsilon$ diverges. We assume that $H_\mathrm{end} = H_\mathrm{reh}$ and $H_\mathrm{end} = H_\mathrm{asymp}$, where the latter quantity can be computed  analytically for the $\{ \epsilon, \eta, \xi\}$ parameterization by solving a quadratic equation to find the zero of $\epsilon$. Then we substitute this $H_\mathrm{end}$ into equation (\ref{lnconnect}) to obtain $N_\mathrm{end}$. The eternal inflation prior is then implemented by requiring that $P_\mathcal{R}(N_\mathrm{end}) \leq 1$. Figure \ref{fig:fr3ei} shows the effect of applying this prior.  We see that the WMAP5+SNLS constraints with no eternal inflation prior are roughly the same as the WMAP5 data combination {\sl  with} this prior - i.e., WMAP5+SNLS is able to rule out the eternally-inflating part of the parameter space which is allowed by WMAP5 alone. 

\section{Discussion}

We present  constraints on the slow roll parameters, following the WMAP5 data release  \cite{Hinshaw:2008kr, Nolta:2008ih, Dunkley:2008ie, Komatsu:2008hk}, in combination with ACBAR  (ACBAR) \cite{Reichardt:2008ay}, and SNLS \cite{Astier:2005qq}.   This analysis updates our treatment of the WMAP3 \cite{Peiris:2006sj,Peiris:2006ug} release.   In addition, we have presented a more careful implementation of the ``e-fold prior'', which we use to eliminate models that do not produce sufficient  inflation.  Further, we have added a prior on the inflationary parameter space that excludes models which  over-produce primordial black holes, or would permit the onset of eternal inflation  after CMB scales have left the horizon.  

We can easily post-process our chains to present our results in terms of the usual $\{n_s,r,dn_s/d\ln{k}\}$ formalism.  With two slow roll parameters, the permitted ranges of $n_s$ and $r$ closely match those obtained by including these parameters directly in the chains.  Conversely, with three slow roll parameters, Slow Roll Reconstruction leads to very tight limits on $dn_s/d\ln{k}$ -- especially when we include a primordial black hole constraint.   It is important to realize that these limits are being driven by our priors, rather than the data. The ability of Slow Roll Reconstruction to include these priors -- especially when they relate  to the duration of inflation -- is a key advantage of this methodology. On the other hand, despite the great strides that have been made in recent years, current cosmological data does not significantly constrain $\xi$, or $V'''$, given that the bounds on this parameter are largely functions of the priors.  We note that the precise form of the primordial black hole bound is currently somewhat ambiguous, and a thorough analysis of this constraint should be undertaken in the future.  However, even with current data, the combination of WMAP5 and SNLS essentially eliminates the region of parameter space that would support eternal inflation. Likewise, with weak assumptions about reheating, the region of parameter space which leads to the overproduction of black holes is roughly similar to that excluded by WMAP5+SNLS. However, if we assume the reheating scale is close to the GUT scale, the prior rules out regions of parameter space currently allowed by the data.

One can view the overlap between the region of the $\{n_s, dn_s/d\ln{k} \}$ plane permitted by a conventional analysis and Slow Roll Reconstruction shown in the lower panel of Figure~\ref{fig:fr3b} as a consistency check -- if these two regions did {\em not} overlap, we could exclude all 3-parameter models of single field inflation. Both eternal inflation and the overproduction of primordial black holes can be avoided by including a fourth slow roll parameter, which makes $\xi$ scale dependent. We do not explore this option here: since $\xi$ is only weakly constrained by the data, the bounds on a fourth slow roll parameter (${}^3\lambda$) would be almost entirely set by the priors.   Further, given that these parameters are not strongly constrained by the data, the central values we compute will be functions of our prior distributions.    Since $\epsilon$ effectively measures the inflationary scale (once the amplitude of the scalar perturbation spectrum is known), a flat prior on $\epsilon$ biases us toward high inflationary scales, whereas a log prior would tip us towards inflationary scales that are that are below the GUT scale -- with the lower bound coming from our assumptions about reheating, as explained in Section~2.    Conversely, different slow roll parameterizations can suggest a ``detection'' of tensor models within current data \cite{Valkenburg:2008cz}.  This ambiguity is a reminder of the importance of priors in Bayesian statistics \cite{Trotta:2008qt} and is characteristic of  a parameter which is not (currently) well-constrained by data.

Needless to say, we expect this situation to improve in the future, but given that the permitted region of the    $\{n_s, dn_s/d\ln{k} \}$ plane is currently much larger than that consistent with the slow roll prior, we need more than incremental improvements in the global cosmological dataset.  It is clear that {\em Planck\/} will lead to a substantial tightening of constraints on the running \cite{Verde:2005ff,:2006uk}. Further, if dedicated polarization experiments detect primordial tensor modes this would put a {\em lower\/} bound on $\epsilon$, removing most of the ambiguity in the inflationary scale, and thus the e-fold and reheating priors built into Slow Roll Reconstruction. Moreover, we can expect similar improvements from large scale structure measurements,  supernovae, or measurements of the primordial spectrum from high redshift 21~cm experiments.  In particular, if $\epsilon$ is not vanishingly small, second order terms in slow roll naturally lead to a small running of the scalar spectral index. This running cannot be detected in CMB data alone \cite{Adshead:2008vn}, but may well be detectable in combinations of several datasets, providing further constraints on the inflationary parameter space.  

 \ack We thank Peter Adshead, Laila Alabidi, Simeon Bird, George Efstathiou, Jan Hamann, Eiichiro Komatsu, Julien Lesgourgues, Jim Lidsey, Karim Malik, and Wessel Valkenburg for useful discussions. HVP is supported in part by Marie Curie grant MIRG-CT-2007-203314 from the European Commission, and by an STFC Advanced Fellowship. RE is supported in part by the United States Department of Energy, grant DE-FG02-92ER-40704 and by an NSF Career Award PHY-0747868. Our numerical computations were performed using the Darwin Supercomputer of the University of Cambridge High Performance Computing Service (http://www.hpc.cam.ac.uk/), provided by Dell Inc. using Strategic Research Infrastructure Funding from the Higher Education Funding Council for England. We acknowledge the use of the Legacy Archive for Microwave Background Data (LAMBDA). Support for LAMBDA is provided by  the NASA Office of Space Science.

\section*{References}
\bibliographystyle{h-physrev.bst}
\bibliography{flowroll}

\begin{thebibliography}{10}

\bibitem{Copeland:1993jj}
E.~J. Copeland, E.~W. Kolb, A.~R. Liddle, and J.~E. Lidsey,
\newblock Phys. Rev. {\bf D48}, 2529 (1993), hep-ph/9303288.

\bibitem{Copeland:1993ie}
E.~J. Copeland, E.~W. Kolb, A.~R. Liddle, and J.~E. Lidsey,
\newblock Phys. Rev. Lett. {\bf 71}, 219 (1993), hep-ph/9304228.

\bibitem{Turner:1993su}
M.~S. Turner,
\newblock Phys. Rev. {\bf D48}, 5539 (1993), astro-ph/9307035.

\bibitem{Liddle:1994cr}
A.~R. Liddle and M.~S. Turner,
\newblock Phys. Rev. {\bf D50}, 758 (1994), astro-ph/9402021.

\bibitem{Peiris:2006sj}
H.~Peiris and R.~Easther,
\newblock JCAP {\bf 0610}, 017 (2006), astro-ph/0609003.

\bibitem{Peiris:2006ug}
H.~Peiris and R.~Easther,
\newblock JCAP {\bf 0607}, 002 (2006), astro-ph/0603587.

\bibitem{Easther:2006tv}
R.~Easther and H.~Peiris,
\newblock JCAP {\bf 0609}, 010 (2006), astro-ph/0604214.

\bibitem{Adshead:2008vn}
P.~Adshead and R.~Easther,
\newblock (2008), 0802.3898.

\bibitem{Hoffman:2000ue}
M.~B. Hoffman and M.~S. Turner,
\newblock Phys. Rev. {\bf D64}, 023506 (2001), astro-ph/0006321.

\bibitem{Kinney:2002qn}
W.~H. Kinney,
\newblock Phys. Rev. {\bf D66}, 083508 (2002), astro-ph/0206032.

\bibitem{Easther:2002rw}
R.~Easther and W.~H. Kinney,
\newblock Phys. Rev. {\bf D67}, 043511 (2003), astro-ph/0210345.

\bibitem{Kinney:2006qm}
W.~H. Kinney, E.~W. Kolb, A.~Melchiorri, and A.~Riotto,
\newblock Phys. Rev. {\bf D74}, 023502 (2006), astro-ph/0605338.

\bibitem{Powell:2007gu}
B.~A. Powell and W.~H. Kinney,
\newblock JCAP {\bf 0708}, 006 (2007), 0706.1982.

\bibitem{Liddle:2003py}
A.~R. Liddle,
\newblock Phys. Rev. {\bf D68}, 103504 (2003), astro-ph/0307286.

\bibitem{Grivell:1996sr}
I.~J. Grivell and A.~R. Liddle,
\newblock Phys. Rev. {\bf D54}, 7191 (1996), astro-ph/9607096.

\bibitem{Lesgourgues:2007aa}
J.~Lesgourgues, A.~A. Starobinsky, and W.~Valkenburg,
\newblock JCAP {\bf 0801}, 010 (2008), 0710.1630.

\bibitem{Hamann:2008pb}
J.~Hamann, J.~Lesgourgues, and W.~Valkenburg,
\newblock JCAP {\bf 0804}, 016 (2008), 0802.0505.

\bibitem{Stewart:1993bc}
E.~D. Stewart and D.~H. Lyth,
\newblock Phys. Lett. {\bf B302}, 171 (1993), gr-qc/9302019.

\bibitem{Leach:2002ar}
S.~M. Leach, A.~R. Liddle, J.~Martin, and D.~J. Schwarz,
\newblock Phys. Rev. {\bf D66}, 023515 (2002), astro-ph/0202094.

\bibitem{Leach:2002dw}
S.~M. Leach and A.~R. Liddle,
\newblock Mon. Not. Roy. Astron. Soc. {\bf 341}, 1151 (2003), astro-ph/0207213.

\bibitem{Leach:2003us}
S.~M. Leach and A.~R. Liddle,
\newblock Phys. Rev. {\bf D68}, 123508 (2003), astro-ph/0306305.

\bibitem{Hinshaw:2008kr}
WMAP, G.~Hinshaw {\em et~al.},
\newblock (2008), 0803.0732.

\bibitem{Nolta:2008ih}
WMAP, M.~R. Nolta {\em et~al.},
\newblock (2008), 0803.0593.

\bibitem{Dunkley:2008ie}
WMAP, J.~Dunkley {\em et~al.},
\newblock (2008), 0803.0586.

\bibitem{Komatsu:2008hk}
WMAP, E.~Komatsu {\em et~al.},
\newblock (2008), 0803.0547.

\bibitem{Reichardt:2008ay}
C.~L. Reichardt {\em et~al.},
\newblock (2008), 0801.1491.

\bibitem{Astier:2005qq}
The SNLS, P.~Astier {\em et~al.},
\newblock Astron. Astrophys. {\bf 447}, 31 (2006), astro-ph/0510447.

\bibitem{Hawking:1971ei}
S.~Hawking,
\newblock Mon. Not. Roy. Astron. Soc. {\bf 152}, 75 (1971).

\bibitem{Carr:1974nx}
B.~J. Carr and S.~W. Hawking,
\newblock Mon. Not. Roy. Astron. Soc. {\bf 168}, 399 (1974).

\bibitem{Green:1997sz}
A.~M. Green and A.~R. Liddle,
\newblock Phys. Rev. {\bf D56}, 6166 (1997), astro-ph/9704251.

\bibitem{Yokoyama:1999xi}
J.~Yokoyama,
\newblock Prog. Theor. Phys. Suppl. {\bf 136}, 338 (1999).

\bibitem{Leach:2000ea}
S.~M. Leach, I.~J. Grivell, and A.~R. Liddle,
\newblock Phys. Rev. {\bf D62}, 043516 (2000), astro-ph/0004296.

\bibitem{Chongchitnan:2006wx}
S.~Chongchitnan and G.~Efstathiou,
\newblock JCAP {\bf 0701}, 011 (2007), astro-ph/0611818.

\bibitem{Zaballa:2006kh}
I.~Zaballa, A.~M. Green, K.~A. Malik, and M.~Sasaki,
\newblock JCAP {\bf 0703}, 010 (2007), astro-ph/0612379.

\bibitem{Kohri:2007qn}
K.~Kohri, D.~H. Lyth, and A.~Melchiorri,
\newblock JCAP {\bf 0804}, 038 (2008), 0711.5006.

\bibitem{Linde:1986fd}
A.~D. Linde,
\newblock Phys. Lett. {\bf B175}, 395 (1986).

\bibitem{Lidsey:1995np}
J.~E. Lidsey {\em et~al.},
\newblock Rev. Mod. Phys. {\bf 69}, 373 (1997), astro-ph/9508078.

\bibitem{Grishchuk:1988}
L.~Grishchuk and Y.~V. Sidorav,
\newblock Fourth Seminar on Quantum Gravity, World Scientific, Singapore.

\bibitem{Muslimov:1990be}
A.~G. Muslimov,
\newblock Class. Quant. Grav. {\bf 7}, 231 (1990).

\bibitem{Salopek:1990jq}
D.~S. Salopek and J.~R. Bond,
\newblock Phys. Rev. {\bf D42}, 3936 (1990).

\bibitem{Salopek:1990re}
D.~S. Salopek and J.~R. Bond,
\newblock Phys. Rev. {\bf D43}, 1005 (1991).

\bibitem{Lidsey:1991zp}
J.~E. Lidsey,
\newblock Phys. Lett. {\bf B273}, 42 (1991).

\bibitem{Adams:2001vc}
J.~A. Adams, B.~Cresswell, and R.~Easther,
\newblock Phys. Rev. {\bf D64}, 123514 (2001), astro-ph/0102236.

\bibitem{Cortes:2007ak}
M.~Cortes, A.~R. Liddle, and P.~Mukherjee,
\newblock Phys. Rev. {\bf D75}, 083520 (2007), astro-ph/0702170.

\bibitem{Dodelson:2003ft}
S.~Dodelson,
\newblock {\em {Modern cosmology}} ,
\newblock Amsterdam, Netherlands: Academic Pr. (2003) 440 p.

\bibitem{Liddle:2003as}
A.~R. Liddle and S.~M. Leach,
\newblock Phys. Rev. {\bf D68}, 103503 (2003), astro-ph/0305263.

\bibitem{Lewis:1999bs}
A.~Lewis, A.~Challinor, and A.~Lasenby,
\newblock Astrophys. J. {\bf 538}, 473 (2000), astro-ph/9911177.

\bibitem{Olive:1999ij}
K.~A. Olive, G.~Steigman, and T.~P. Walker,
\newblock Phys. Rept. {\bf 333}, 389 (2000), astro-ph/9905320.

\bibitem{Lyth:1995ka}
D.~H. Lyth and E.~D. Stewart,
\newblock Phys. Rev. {\bf D53}, 1784 (1996), hep-ph/9510204.

\bibitem{Burgess:2005sb}
C.~P. Burgess, R.~Easther, A.~Mazumdar, D.~F. Mota, and T.~Multamaki,
\newblock JHEP {\bf 05}, 067 (2005), hep-th/0501125.

\bibitem{Chung:2007vz}
D.~J.~H. Chung, L.~L. Everett, and K.~T. Matchev,
\newblock Phys. Rev. {\bf D76}, 103530 (2007), 0704.3285.

\bibitem{Lewis:2002ah}
A.~Lewis and S.~Bridle,
\newblock Phys. Rev. {\bf D66}, 103511 (2002), astro-ph/0205436.

\bibitem{Gelman92}
A.~Gelman and D.~Rubin,
\newblock Statistical Science {\bf 7}, 457 (1992).

\bibitem{GarciaBellido:1996qt}
J.~Garcia-Bellido, A.~D. Linde, and D.~Wands,
\newblock Phys. Rev. {\bf D54}, 6040 (1996), astro-ph/9605094.

\bibitem{Stewart:1996ey}
E.~D. Stewart,
\newblock Phys. Lett. {\bf B391}, 34 (1997), hep-ph/9606241.

\bibitem{Stewart:1997wg}
E.~D. Stewart,
\newblock Phys. Rev. {\bf D56}, 2019 (1997), hep-ph/9703232.

\bibitem{Covi:1998jp}
L.~Covi, D.~H. Lyth, and L.~Roszkowski,
\newblock Phys. Rev. {\bf D60}, 023509 (1999), hep-ph/9809310.

\bibitem{Covi:1998mb}
L.~Covi and D.~H. Lyth,
\newblock Phys. Rev. {\bf D59}, 063515 (1999), hep-ph/9809562.

\bibitem{Covi:2002th}
L.~Covi, D.~H. Lyth, and A.~Melchiorri,
\newblock Phys. Rev. {\bf D67}, 043507 (2003), hep-ph/0210395.

\bibitem{Covi:2004tp}
L.~Covi, D.~H. Lyth, A.~Melchiorri, and C.~J. Odman,
\newblock Phys. Rev. {\bf D70}, 123521 (2004), astro-ph/0408129.

\bibitem{Carr:1994ar}
B.~J. Carr, J.~H. Gilbert, and J.~E. Lidsey,
\newblock Phys. Rev. {\bf D50}, 4853 (1994), astro-ph/9405027.

\bibitem{Kim:1996hr}
H.~I. Kim and C.~H. Lee,
\newblock Phys. Rev. {\bf D54}, 6001 (1996).

\bibitem{Green:1999yh}
A.~M. Green,
\newblock Phys. Rev. {\bf D60}, 063516 (1999), astro-ph/9903484.

\bibitem{Lemoine:2000sq}
M.~Lemoine,
\newblock Phys. Lett. {\bf B481}, 333 (2000), hep-ph/0001238.

\bibitem{Valkenburg:2008cz}
W.~Valkenburg, L.~M. Krauss, and J.~Hamann,
\newblock (2008), 0804.3390.

\bibitem{Trotta:2008qt}
R.~Trotta,
\newblock (2008), 0803.4089.

\bibitem{Verde:2005ff}
L.~Verde, H.~Peiris, and R.~Jimenez,
\newblock JCAP {\bf 0601}, 019 (2006), astro-ph/0506036.

\bibitem{:2006uk}
Planck,
\newblock (2006), astro-ph/0604069.

\end{thebibliography}
\newpage
\appendix
\section{Parameter Constraints  from Monte-Carlo Markov Chain Analysis \label{appendix}}

\begin{table}[!th]{
\begin{tabular}{||c|c|c|c||}
\hline
$T_\mathrm{reh}> 10\ \mathrm{MeV}$          & WMAP5   	   &  WMAP5 + ACBAR           &  WMAP5 + SNLS \\
\hline 
\hline 
$\epsilon$          &  $<0.020$ (95\% CL)  &   $<0.018$ (95\% CL)  &  $<0.016$ (95\% CL) \\
\hline 
$\eta$      	        &  $ 0.009_{-0.021}^{+0.022}$  &   $ 0.007_{-0.018}^{+0.020}$  &  $ 0.002_{-0.016}^{+0.016}$ \\
\hline
$\ln[10^{10} A_s]$      &   $  3.07_{-0.04}^{+0.04}$ &  $  3.08_{-0.04}^{+0.04}$  &  $  3.08_{-0.04}^{+0.04}$ \\
\hline
\multicolumn{4}{c}{}\\
\hline
$T_\mathrm{reh}> 10\ \mathrm{TeV}$          & WMAP5   	   &  WMAP5 + ACBAR           &  WMAP5 + SNLS \\
\hline 
\hline 
$\epsilon$              &    $<0.020$ (95\% CL) &   $<0.018$ (95\% CL) &  $<0.016$ (95\% CL) \\
\hline 
$\eta$      	        &   $ 0.009_{-0.020}^{+0.021}$ &  $ 0.008_{-0.019}^{+0.020}$  & $ 0.002_{-0.016}^{+0.017}$  \\
\hline
$\ln[10^{10} A_s]$      &   $  3.07_{-0.04}^{+0.04}$  &  $  3.08_{-0.04}^{+0.04}$ &  $  3.08_{-0.04}^{+0.04}$ \\
\hline
\multicolumn{4}{c}{}\\
\hline
$H_\mathrm{end} = H_\mathrm{reh} $          & WMAP5   	   &  WMAP5 + ACBAR           &  WMAP5 + SNLS \\
\hline 
\hline 
$\epsilon$              &  $<0.020$ (95\% CL)  &   $<0.018$ (95\% CL) & $<0.016$ (95\% CL)  \\
\hline 
$\eta$      	        &   $ 0.012_{-0.021}^{+0.021}$ &  $ 0.009_{-0.018}^{+0.019}$ &  $ 0.005_{-0.017}^{+0.016}$ \\
\hline
$\ln[10^{10} A_s]$      &   $  3.07_{-0.04}^{+0.04}$ & $  3.08_{-0.04}^{+0.04}$  &  $  3.08_{-0.04}^{+0.04}$ \\
\hline
\end{tabular}
\caption{Constraints on HSR parameters defined at $k=0.02$ Mpc$^{-1}$ for the $\{\epsilon, \eta, \log [10^{10} A_s] \}$ parameterization. Constraints are at the 68\% confidence level unless otherwise noted.}\label{table:hsr_constraints_fr2}}
\end{table}

\begin{table}[!th]{
\begin{tabular}{||c|c|c|c||}
\hline
$T_\mathrm{reh}> 10\ \mathrm{MeV}$          & WMAP5   	   &  WMAP5 + ACBAR           &  WMAP5 + SNLS \\
\hline 
\hline 
$\epsilon$              &   $<0.020$ (95\% CL) &   $<0.018$ (95\% CL) & $<0.016$ (95\% CL) \\
\hline 
$\eta$      	        &   $ 0.017_{-0.023}^{+0.023}$ & $ 0.012_{-0.019}^{+0.019}$  &  $ 0.007_{-0.017}^{+0.017}$ \\
\hline
$\xi$      	        &  $-0.007_{-0.007}^{+0.006}$  &  $-0.005_{-0.005}^{+0.005}$ &  $-0.005_{-0.005}^{+0.005}$ \\
\hline
$\ln[10^{10} A_s]$      & $  3.05_{-0.05}^{+0.05}$   &  $  3.06_{-0.04}^{+0.04}$ & $  3.07_{-0.04}^{+0.04}$  \\
\hline
\multicolumn{4}{c}{}\\
\hline
 $T_\mathrm{reh}> 10\ \mathrm{TeV}$          & WMAP5   	   &  WMAP5 + ACBAR           &  WMAP5 + SNLS \\
\hline 
\hline 
$\epsilon$              &  $<0.020$ (95\% CL)  &  $<0.018$ (95\% CL) & $<0.015$ (95\% CL)  \\
\hline 
$\eta$      	        & $ 0.018_{-0.022}^{+0.023}$   &  $ 0.012_{-0.019}^{+0.020}$ &  $ 0.007_{-0.016}^{+0.017}$ \\
\hline
$\xi$      	        &  $-0.007_{-0.007}^{+0.006}$  & $-0.005_{-0.005}^{+0.005}$  & $-0.006_{-0.005}^{+0.005}$  \\
\hline
$\ln[10^{10} A_s]$      &   $  3.05_{-0.05}^{+0.05}$ & $  3.07_{-0.04}^{+0.04}$ & $  3.07_{-0.04}^{+0.04}$  \\
\hline
\multicolumn{4}{c}{}\\
\hline
$H_\mathrm{end} = H_\mathrm{reh} $          & WMAP5   	   &  WMAP5 + ACBAR           &  WMAP5 + SNLS \\
\hline 
\hline 
$\epsilon$              &   $<0.020$ (95\% CL) &  $<0.017$ (95\% CL)  &  $<0.015$ (95\% CL) \\
\hline 
$\eta$      	        & $ 0.017_{-0.022}^{+0.022}$   & $ 0.012_{-0.019}^{+0.019}$  & $ 0.006_{-0.016}^{+0.017}$  \\
\hline
$\xi$      	        &  $-0.007_{-0.006}^{+0.006}$  & $-0.005_{-0.005}^{+0.005}$  &  $-0.006_{-0.005}^{+0.005}$ \\
\hline
$\ln[10^{10} A_s]$      &   $  3.05_{-0.05}^{+0.05}$ &  $  3.07_{-0.04}^{+0.04}$ & $  3.07_{-0.04}^{+0.04}$  \\
\hline
\end{tabular}
\caption{Constraints on HSR parameters  defined at $k=0.02$ Mpc$^{-1}$ for the $\{\epsilon, \eta, \xi, \log [10^{10} A_s] \}$ parameterization. Constraints are at the 68\% confidence level unless otherwise noted.}\label{table:hsr_constraints_fr3}}
\end{table}

\end{document}